\documentstyle[psfig]{mn}
\newcommand\etal{et al. }
\newcommand\refer{\par \noindent\hangindent=3pc \hangafter=1}

\newcommand\ros{{\it ROSAT}}
\newcommand\asca{{\it ASCA}}

\newcommand\eg{eg }

\newcommand\ie{i.e. }
\newcommand\ecs{erg cm$^{-2}$ s$^{-1}$}
\newcommand\nh{N_{H}}
\newcommand\ftools{FTOOLS}

\title{A Survey of hard spectrum \ros\ sources 1: X--ray source catalogue}
\author[Page, Mittaz \& Carrera]{M.J. Page\(^{1}\), J.P.D. Mittaz\(^{1}\), 
F.J. Carrera\(^{1,2}\)\\
\(^{1}\)Mullard Space Science Laboratory, University College London,
Holmbury St Mary, Dorking, Surrey RH5 6NT, UK.\\
\(^{2}\)Instituto de F\'\i sica de Cantabria (Consejo Superior de
Investigaciones Cient\'\i ficas--Universidad de Cantabria), 39005
Santander, Spain.}

\date{}

\begin{document}
\maketitle

\begin{abstract}

We present a catalogue of 147 serendipitous X--ray sources selected to have
hard spectra (\(\alpha < 0.5\)) from a survey of 188 \ros\ fields. Such sources
must be the dominant contributors to the X-ray background at faint fluxes.  We
have used Monte Carlo simulations to verify that our technique is very
efficient at selecting hard sources: the survey has \(\geq\) 10 times as much
effective area to hard sources as it has to soft sources above a 0.5 - 2 keV
flux level of \(10^{-14}\) \ecs.  The distribution of best fit spectral slopes
of the hard sources suggests that a typical \ros\ hard source in our survey
has a spectral slope \(\alpha \sim 0\).  The hard sources have a steep number
flux relation
(\(dN/dS \propto S^{-\gamma}\) with a best fit value of \(\gamma = 2.72\pm
0.12\)) and make up about 15\% of all 0.5 - 2 keV sources with \(S > 10^{-14}\)
\ecs. If their \(N(S)\) continues to fainter fluxes, the hard
sources will comprise \(\sim 40\%\) of sources with \(5 \times 10^{-15}\) \ecs\
\(< S <\) \(10^{-14}\) \ecs. The population of hard sources can therefore
account for the harder average spectra of \ros\ sources with \(S < 10^{-14}\)
\ecs.  They probably make a strong contribution to the X-ray background at
faint fluxes and could be the solution to the X-ray background 
spectral paradox.
 
\end{abstract}

\section{Introduction}

Determining the origin of the extragalactic X--ray background (XRB)
has been a major goal of X--ray astronomers for more than three
decades since its discovery (Giacconi \etal 1962), and surveys of the
soft X--ray sky with {\it Rosat} have succeeded in resolving \(\sim 80
\%\) of the 1-2 keV XRB into individual sources (Hasinger \etal
1998). For the brighter sources which produce \(\sim 40\%\) of the XRB, 
X--ray spectroscopy and optical identification has been
possible.  The majority of these sources are active galactic nuclei
(AGN) with broad emission lines, \ie Seyfert 1 galaxies and QSOs.  At
the faintest fluxes probed, a population of narrow emission line
galaxies (NELGs) has been detected (McHardy \etal 1998). It has been
argued that many of these are also AGN, but with low luminosity or obscured
broad line regions (Schmidt \etal 1998).

However, it is not possible to synthesise the entire XRB by
extrapolating the observed source populations to faint fluxes, because
the majority of the known sources have X--ray spectra which are softer
than the background.  This discrepancy is present for all energy bands
between 0.5 and 40 keV, and the 0.5 - 2 keV band where the deepest
surveys have taken place is no exception.  The spectrum of the
extragalactic background, as measured with past and current
instruments between 1 and 10 keV, can be described by a power law \(F_{\nu}
\propto \nu^{-\alpha}\) where \(\alpha = 0.4 \pm0.1\) (Chen, Fabian \&
Gendreau 1997, Miyaji et al 1998).  The integrated spectrum of the
broad line AGN detected in current 0.5 - 2 keV X--ray surveys is much
softer, a power law with \(\alpha \geq 1\) (Mittaz \etal 1999, Ciliegi \etal 1994), 
while that of the faint 0.5 - 2 keV sources
identified as NELGs is similar to the background: Romero-Colmenero
\etal (1996) found \(\alpha = 0.4 \pm 0.1\) and Almaini \etal (1996)
\(\alpha = 0.5 \pm 0.1\) for two independent X--ray selected NELG
samples.  
The integrated spectrum of these two populations is softer than the spectrum of
the X-ray background, hence much of the remaining background must be produced
by sources with spectra which are harder than the mean spectra of currently
identified NELGs and AGN.  
The composition of the hard source population has not yet been determined,
although we might expect a significant overlap with the NELG population,
 because the fitted X--ray
spectral slopes of individual NELGs show considerable scatter, and because
faint optical galaxies make a significant fraction ($\sim 30-40 \%$) of the 
remaining unresolved X--ray background in deep PSPC images (Roche \etal 1995,
Almaini \etal 1997, Newsam \etal 1999).

Current models for the synthesis of the X--ray background propose that
a large proportion of the remaining background sources are obscured
AGN. Such sources would be expected to have both lower fluxes and
harder spectra than the current source populations. From X-ray
spectroscopy of a large sample of sources from the \ros\ International
X--ray Optical Survey (RIXOS, Mason \etal 2000), Mittaz \etal (1999)
concluded that at faint \ros\ fluxes (\(< 10^{-14}\) \ecs\ ) \(\sim
13\%\) of sources have spectra harder than that of the background,
compared with only \(\sim 7\%\) at brighter fluxes. This is probably
the bright tail of the hard source population that must make a
substantial contribution to the background at even fainter
fluxes. These \ros\ detected hard sources should therefore provide us with a
preview of the XRB producing population. As relatively bright
examples, they are likely to be much easier to study at all wavelengths than
their more numerous, but fainter, cousins.

This paper describes the survey of \ros\ PSPC pointed observations for examples
of this faint, hard spectrum population and presents the catalogue of hard
spectrum sources. Optical identification, spectroscopy and imaging of these
sources are the subjects of companion papers. 
We describe the
construction of the source catalogue and the \ros\ data reduction method in
Section \ref{sec:method}.  The Monte Carlo simulations which were used to
calculate the effective area of the survey and quantify the spectral selection
effects are described in Section \ref{sec:monte}.  The results of the
simulations are taken into account in Section \ref{sec:results} to derive the
characteristic X--ray spectral properties of the hard sources, their source
counts, and their contribution to the 1 - 2 keV XRB. The source catalogue is
also presented in this section.  The implications of our findings are discussed
in Section \ref{sec:discussion}, and our conclusions are presented in Section
\ref{sec:conclusions}.

\section{Construction of the X--ray source catalogue}
\label{sec:method}

\subsection{General description}
\label{sec:description}

The primary goal of the survey is to investigate the properties of
extragalactic sources which have X--ray spectra harder than the
extragalactic background emission, because such sources must exist in
abundance at faint fluxes to resolve the spectral paradox.  We chose
to construct our source catalogue using archival \ros\ PSPC data.
This was well suited to our purposes: the \ros\ PSPC has
sufficient energy resolution to effectively discriminate hard from
soft sources, while the excellent spatial resolution means we can
expect to make unambiguous optical identifications for many of the
hard sources.
 
\subsection{Choice of \ros\ observations}

The \ros\ PSPC fields were chosen according to the following
preferences: high Galactic latitude, low Galactic \(\nh\), long exposure
time, observation targets which did not fill a large fraction of the field of
view, and sky positions suitable for optical follow up.  
Only PSPC observations with the `open' filter
were used, and observations targeted on the Magellanic clouds were
excluded.  Some 188 PSPC datasets were searched for hard sources; a
complete list of these observations is given in table
\ref{tab:fields}.

\begin{table}
\caption{\ros\ observations searched for hard sources.}
\label{tab:fields}
\begin{tabular}{llll}
rp200921n00& 
rp900140n00& 
rp600541n00& 
rp800344n00\\
rp800482n00& 
rp800483n00& 
rp600025a01& 
rp701217a01\\
rp800367a01& 
rp800383n00& 
rp700423n00& 
rp300133n00\\
rp800486n00& 
rp900190n00& 
rp800304n00& 
rp600439n00\\
rp300334n00& 
rp800326n00& 
rp400080n00& 
rp600504n00\\
rp600107n00& 
rp701180a01& 
rp700920n00& 
rp201488n00\\
rp700468n00& 
rp700424a01& 
rp300004n00& 
rp700884n00\\
rp900138n00& 
rp201339n00& 
rp700448n00& 
rp701187n00\\
rp900495n00& 
rp700873n00& 
rp600133n00& 
rp701223n00\\
rp700185n00& 
rp600159a00& 
rp700924a01& 
rp600005n00\\
rp700049n00& 
rp700882n00& 
rp900352n00& 
rp700516a00\\
rp201094n00& 
rp300007a02& 
rp701036n00& 
rp700467n00\\
rp900632n00& 
rp300079n00& 
rp800368n00& 
rp800226n00\\
rp701356n00& 
rp800386n00& 
rp700275n00& 
rp900496a01\\
rp800316n00& 
rp600436n00& 
rp600163n00& 
rp200208n00\\
rp800388n00& 
rp150082n00& 
rp600006n00& 
rp201374a01\\
rp300016n00& 
rp300389n00& 
rp700101n00& 
rp200654n00\\
rp600106n00& 
rp900147n00& 
rp900339n00& 
rp201552n00\\
rp701407n00& 
rp701499n00& 
rp800469n00& 
rp900009a01\\
rp201367m01& 
rp600050n00& 
rp600108n00& 
rp600262a02\\
rp700277n00& 
rp700875n00& 
rp900137n00& 
rp201382n00\\
rp600270n00& 
rp900211n00& 
rp600546n00& 
rp700061n00\\
rp200329n00& 
rp700122n00& 
rp701000a01& 
rp701457n00\\
rp000049n00& 
rp000054n00& 
rp170154n00& 
rp700228n00\\
rp700232n00& 
rp200322n00& 
rp700223n00& 
rp700221n00\\
rp300137n00& 
rp200453n00& 
rp700264n00& 
rp700210n00\\
rp700255n00& 
rp700258n00& 
rp150046n00& 
rp700211n00\\
rp700248n00& 
rp400059n00& 
rp700208n00& 
rp701200n00\\
rp700546n00& 
rp701202n00& 
rp200721n00& 
rp100308n00\\
rp200076n00& 
rp700073n00& 
rp200091n00& 
rp400020n00\\
rp700263n00& 
rp700265a01& 
rp900213n00& 
rp900214n00\\
rp900215n00& 
rp700112n00& 
rp700120n00& 
rp700123n00\\
rp700230a01& 
rp700246n00& 
rp700547n00& 
rp700055n00\\
rp700099m01& 
rp300003n00& 
rp700329a01& 
rp700117n00\\
rp700436n00& 
rp700372n00& 
rp700319n00& 
rp700387n00\\
rp700391n00& 
rp700315n00& 
rp200510n00& 
rp700326n00\\
rp700358n00& 
rp700010n00& 
rp300158n00& 
rp201103n00\\
rp701048n00& 
rp700375n00& 
rp700216a00& 
rp700435a00\\
rp700376n00& 
rp700392n00& 
rp200774n00& 
rp700271n00\\
rp700510n00& 
rp700489n00& 
rp700384n00& 
rp700227n00\\
rp300135n00& 
rp700290n00& 
rp700473n00& 
rp701034n00\\
rp700531n00& 
rp201219n00& 
rp701092n00& 
rp700527n00\\
rp701055n00& 
rp400141n00& 
rp700774n00& 
rp700499n00\\
rp700506n00& 
rp700496n00& 
rp700389n00& 
rp200127a01\\
rp200468n00& 
rp200473n00& 
rp200474n00& 
rp300222n00\\
rp300287n00& 
rp300291n00& 
rp700262a00& 
rp700461n00\\
rp700540n00& 
rp700872n00& 
rp700887n00& 
rp701214n00\\
\end{tabular}
\end{table}

For maximum reliability and reproducibility only REV2 processed data have been
used. For consistency every PSPC dataset was reduced using the same sequence of
operations which will now be described.

\subsection{X--ray data reduction}
\label{sec:reduction}

Each PSPC dataset was first passed through the \ftools\ PCPICOR task
to correct for PSPC spatial/temporal gain variations. The dataset was
then converted to STARLINK ASTERIX format and reduced using the
STARLINK ASTERIX package. The data were screened to remove times of
poor attitude solution, high particle background, and high overall
background countrate.

An image for source searching was then produced. We chose to use the central 20
arcminute radius region where the point spread function and sensitivity are
best, and used only PI channels 19 - 201 so that `ghost imaging' in channels
below 19 would not degrade the positional resolution (Hasinger \etal 1993a,
Snowdon \etal 1994). A mean background level was determined from relatively
source free parts of the image, and the STARLINK PSS source searching routine
was used to find sources more than 4\(\sigma\) above this background.

Next, we proceeded using a method similar to that described in Mittaz \etal
(1999) to derive the spectral parameters of the sources.
Images were constructed in 3 X--ray energy
bands: PI channels 11-41, 52-90 and 91-201. Source counts were
extracted around each PSS source position in each of these 3
images. For most sources a 54 arcsecond radius source circle was used,
corresponding to \(\sim 90\%\) of the counts from a point source. For
sources with one or more contaminating sources nearby, the source
circles were reduced in size to be non-overlapping.  The overall
background estimate for each energy band was obtained by first masking
out all the sources, and then flattening the image by dividing by an
exposure map constructed with the same good time intervals as the
screened data. The expected number of background counts in each source
circle (in each energy band) was then obtained by dividing the counts
in the background image by the unmasked background area, then
multiplying by the area of the source circle and the value of the
exposure map at the position of the source circle. Because the
background count collecting area was much larger than any of the
source circles, the statistical uncertainty on the predicted number of
background counts in the source circle is small compared to the
poisson shot noise on the number of source and background counts
within the source circle.

\subsection{Spectral fitting}
\label{sec:fitting}

To obtain a useful characterization of the X--ray spectrum of each
source, a power law model (\(F_{E} = k E^{-\alpha}\)) was fit to the 3
colour data.  This model has two very useful properties for our study:
it has only two free parameters (slope \(\alpha\) and normalization k)
leaving one degree of freedom when fitting the 3 colour data, and its
shape can easily be compared to that of the extragalactic XRB which is
well described by a power law model (see Section
\ref{sec:criteria}).

The fitting was performed using the method developed by Mittaz \etal
(1999), which is based on the Cash statistic (Cash 1979), a maximum
likelihood estimator appropriate for the Poisson regime.  In this
method, the best fit values for the parameters \(\alpha\) and \(k\) are
obtained by minimising the quantity:
\begin{equation}
C'=-2\sum^{3}_{i=1}obs_{i}\ log(pred_{i}) - pred_{i}
\label{eq:cash}
\end{equation}
where subscript \(i\) denotes the X--ray energy band, \(obs_{i}\) is
the observed (source + background) counts within the source circle and
\(pred_{i}\) is the model predicted, (source + background) counts
within the source circle, given by:
\begin{equation}
pred_{i}=PSF_{i}\times model_{i}(\alpha,k,\nh) + b_{i}
\label{eq:pred}
\end{equation}
\(PSF_{i}\) is the fraction of the point spread function contained
within the source circle for energy band \(i\) calculated using the
equations of Hasinger \etal (1994). \(model_{i}\) is the model
predicted source counts in energy band \(i\) for a power law of slope
\(\alpha\), normalisation \(k\), absorbed by Galactic \(\nh\), obtained
using the \ftools\ `nh' program to interpolate the data of Dickey \&
Lockman (1990).  

For each source a grid of \(\Delta C'\) was used to generate joint
confidence intervals for the fit parameters \(\alpha\) and \(k\),
exactly as \(\Delta \chi^{2}\) is used to produce confidence intervals
in standard \(\chi^{2}\) fitting. It was desirable to obtain from each
two dimensional confidence region a one dimensional confidence
interval in \(\alpha\) for our spectral selection criterion, and a one
dimensional confidence interval in broadband (0.5 - 2.0 keV) flux, the
standard flux measure in \ros\ PSPC surveys. Because these confidence
contours are non-symmetric in many cases, one dimensional marginalised
errors for \(\alpha\) were obtained by integrating the two dimensional
probability distributions along the \(k\) dimension (Loredo 1990 and
references therein).  Similarly, after transforming the probability
distribution from (\(\alpha\),\(k\)) space to (\(\alpha\),Flux)
marginalised errors on the Flux were obtained by integrating over the
\(\alpha\) dimension.

We refer the reader to Section 5 of Mittaz \etal (1999) for derivation
of equations \ref{eq:cash} and \ref{eq:pred} and detailed discussion
of the advantages of this method for fitting faint \ros\ source
spectra. Our application of this method as a sample selection tool is
discussed in Section \ref{sec:effarea}.

\subsection{Radial profile fitting}
\label{sec:radial}

For each source detected by PSS in the channel 19-201 image, we constructed a
radial profile with 5 arcsecond radial bins. We then fitted a model \ros\ PSPC
point spread function appropriate for the source offaxis angle (Hasinger \etal
1993a) to the radial profile using \(\chi^{2}\) out to 1 arcminute.  This was
not used for selecting hard sources, but the value of the best fit
\(\chi^{2}/\nu\) is a useful indicator as to whether a source is point-like,
and is given in column 7 of Table \ref{tab:hardcat} for each hard source.

\section{Criteria for inclusion in the catalogue}
\label{sec:criteria}

A source with
spectral slope \(\alpha^{+u}_{-l}\) is included in the sample if
\begin{equation}
\alpha+u < 0.5
\label{eq:spectralcriterion}
\end{equation}
In other words, the spectral criterion by which a source merits inclusion in 
our hard
source sample is to have the entire of its 68\% best fit confidence
interval in \(\alpha\) harder than the slope of the extragalactic
background, which we take to be \(\alpha=0.5\). 

A spectrum which satisfies this criterion would be expected to have a
probability of \(\alpha < 0.5\) of at least 84\%; in practive most of
the sources included are harder than this at a much higher level of
confidence. For each source we have calculated the probability that
its spectrum is harder than \(\alpha=0.5\) by integrating the
marginalised one dimensional probability distribution in \(\alpha\) up
to \(\alpha=0.5\).  This probability is given for each source in
column 16 of Table \ref{tab:hardcat}.

Finally, in order to produce an unbiased sample of hard sources we rejected any
hard sources which are, or are related to, the target of the PSPC
observation in which they were found. For example, a number of the
observation targets were bright optical galaxies and hard sources
found within these optical galaxies were rejected, because they are probably 
X-ray sources within the galaxies.

\section{Monte Carlo simulations}
\label{sec:monte}

\subsection{Effective area of the survey}
\label{sec:effarea}

The hard source sample is not `flux limited' in the usual style of
X-ray surveys. Indeed, the use of our hard spectral selection
criterion means that the probability of a source at a given flux and
with a given spectrum being included in the sample depends on 1)
Galactic \(\nh\), 2) exposure time, 3) background intensity and
spectrum, 4) local source density (which may affect the size of the
source extraction circle), 5) offaxis angle and of course 6) the
source spectrum. Because items 1 - 3 differ considerably between PSPC
observations, while number 6 is different for every source, imposing
meaningful flux limits is impossible.

Instead we have estimated the total effective area of the survey,
as a function of flux and spectrum, using Monte Carlo simulations.
First, we calculated the geometric area of each field excluding the
area covered by the observation target (see Section
\ref{sec:criteria}).  Then, for every field we simulated 3 colour
spectra of 4000 sources, with predetermined values of flux and
spectral slope. All the other inputs to the simulation were chosen to
reflect the real survey as far as possible. Exposure time and Galactic
\(\nh\) were fixed at the values for the real PSPC field. Source
offaxis angle was generated randomly.  Each source was simulated with
the background found in the real PSPC field and using the real
exposure map.  The source extraction circle size was taken to be that
of the real source most similar in offaxis angle to the simulated
source.  These simulated 3 colour spectra were then fit with a power
law model exactly as for the real sources, and tested with the
spectral selection criterion given in Eq. \ref{eq:spectralcriterion}.
The fraction of simulated sources which pass this selection criterion
is equivalent to the probability of a source in that field, with the
input flux and spectral slope, being included in the hard survey. The
effective area of the field, to sources of that flux
and spectral slope, is therefore the fraction of simulated sources which
pass Eq. \ref{eq:spectralcriterion} multiplied by the geometric area of
the field. The total effective area of our survey is the sum of
the effective areas of all the fields. Note that simulations for
every field were
included in the
effective area calculation, including fields that did not contribute any real 
hard sources
to the survey.

\begin{figure*}
\begin{center}
\leavevmode
\psfig{figure=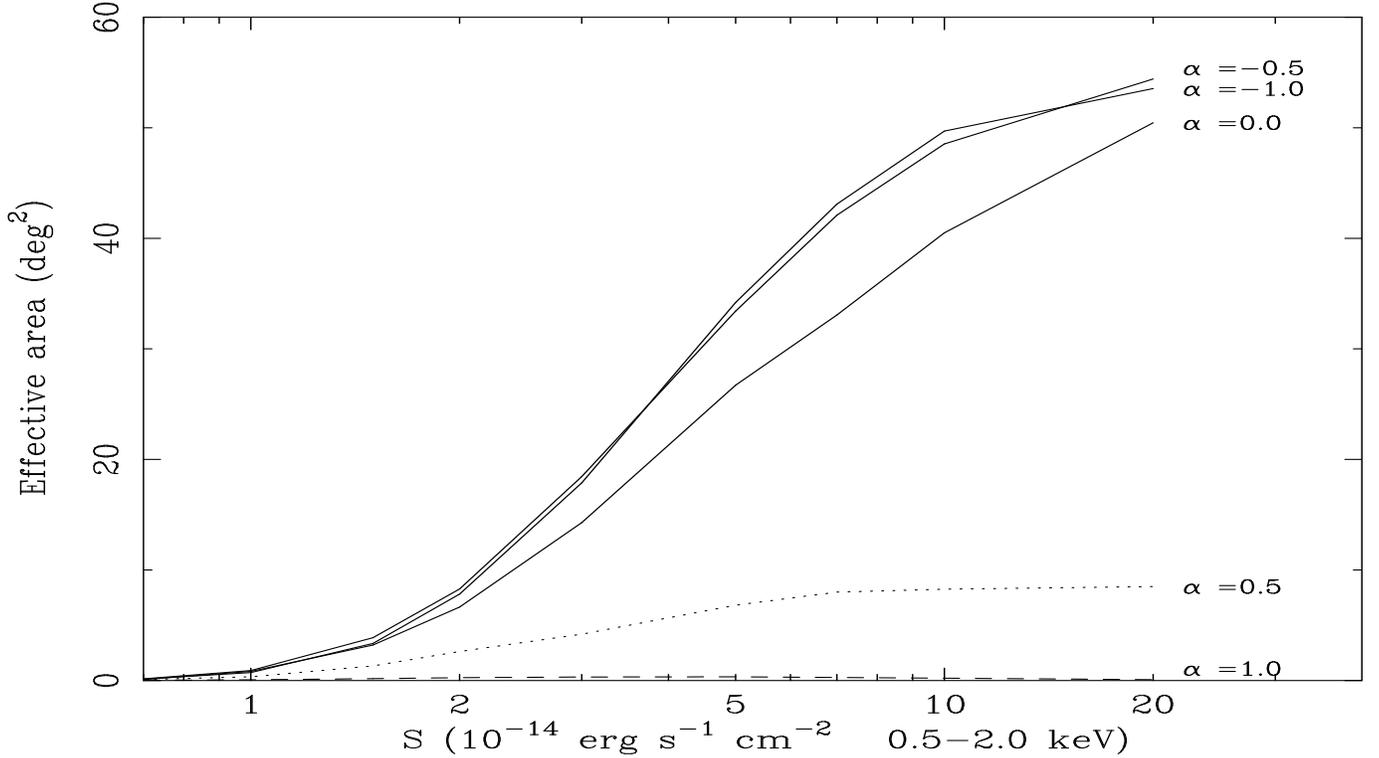,height=100truemm,width=180truemm,angle=270}
\caption{Effective area of the survey as a function of flux for sources with
different power law spectral slopes $\alpha$.}
\label{fig:effareas}
\end{center}
\end{figure*}

The simulations were performed with a grid of input fluxes and spectral slopes
to produce the effective area curves for different spectral slopes given in
Fig. \ref{fig:effareas}. The simulations show for flux \(S > 10^{-14}\) \ecs\
the survey
effective area for hard sources (\(\alpha=0.0\), \(\alpha=-0.5\), or
\(\alpha=-1.0\), solid lines in Fig. \ref{fig:effareas}) 
is at least 10 times larger than the effective area for
typical (\(\alpha=1.0\)) sources (dashed line in Fig. \ref{fig:effareas}).

\subsection{Recovery of source spectra and fluxes}
\label{sec:recovery}

Our spectral selection criterion will inevitably lead to some systematic bias
between the actual and fitted fluxes and spectra of the sources which are
selected.  The simulations used to determine the effective area also allow us
to investigate the relationship between sources' intrinsic spectra and fluxes,
and those obtained after the fitting and selection procedure. This is important
for reconstructing the \(N(S)\) relation and for inferring the spectra of
the hard sources.  In this section, and throughout the rest of the paper, we
use \(F\) to
refer to the output fitted source flux, and \(S\) to refer to a source's input
(intrinsic) flux, so the ratio of fitted to intrinsic flux is \(F/S\).

\begin{figure*}
\begin{center}
\leavevmode
\psfig{figure=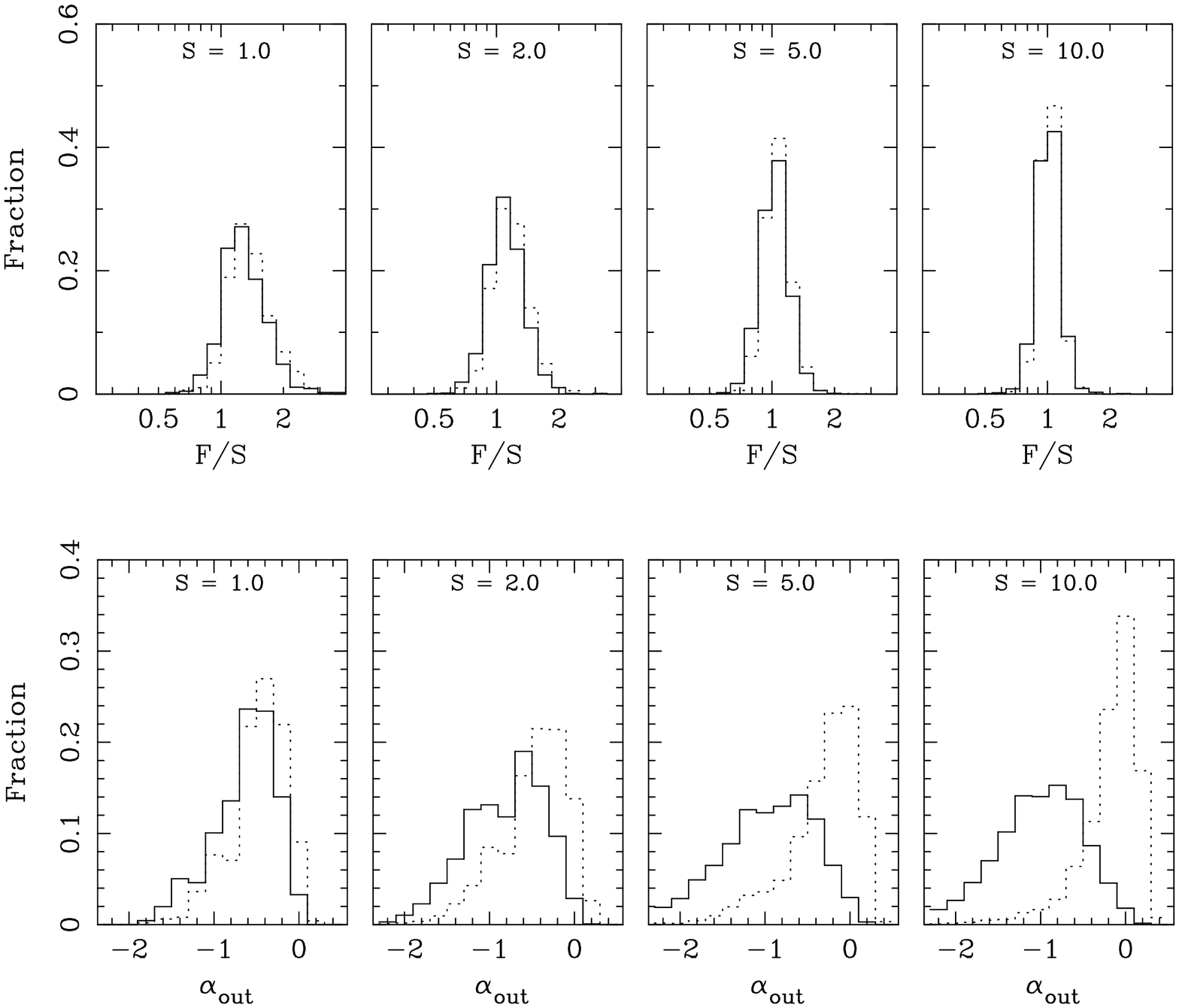,width=170truemm,angle=270}
\caption{Distributions of $F/S$ (top panel, where $S$ is the intrinsic flux and
$F$
is the fitted flux) and fitted spectral slope (bottom panel) for simulated hard
sources with intrinsic $\alpha = -1$ (solid histograms) and $\alpha = 0$
(dashed histograms) which achieve the spectral selection criterion
(Eq. \ref{eq:spectralcriterion}). Note that each histogram is normalised, and
does not represent the sensitivity of the survey to those sources, 
which is given by Fig. \ref{fig:effareas}.}
\label{fig:twohard}
\end{center}
\end{figure*}

The distributions of fitted spectral slopes and \(F/S\) ratios of
hard sources with input spectral slopes of \(\alpha = -1\) (solid histogram)
and \(\alpha = 0\) (dashed histogram) are shown in Fig. \ref{fig:twohard}. For
\(S> 2 \times 10^{-14}\) \ecs\ the input fluxes of the
sources are well recovered (\(\pm 20\%\)) and the peaks of the distributions of
output spectra are close to the input slopes. At fainter fluxes, the output
fluxes are skewed to higher values than the input fluxes, and the distributions
of output spectral slopes are almost indistinguishable, with almost all sources
having a fitted \(\alpha\) of \(< 0\).

\begin{figure*}
\begin{center}
\leavevmode
\psfig{figure=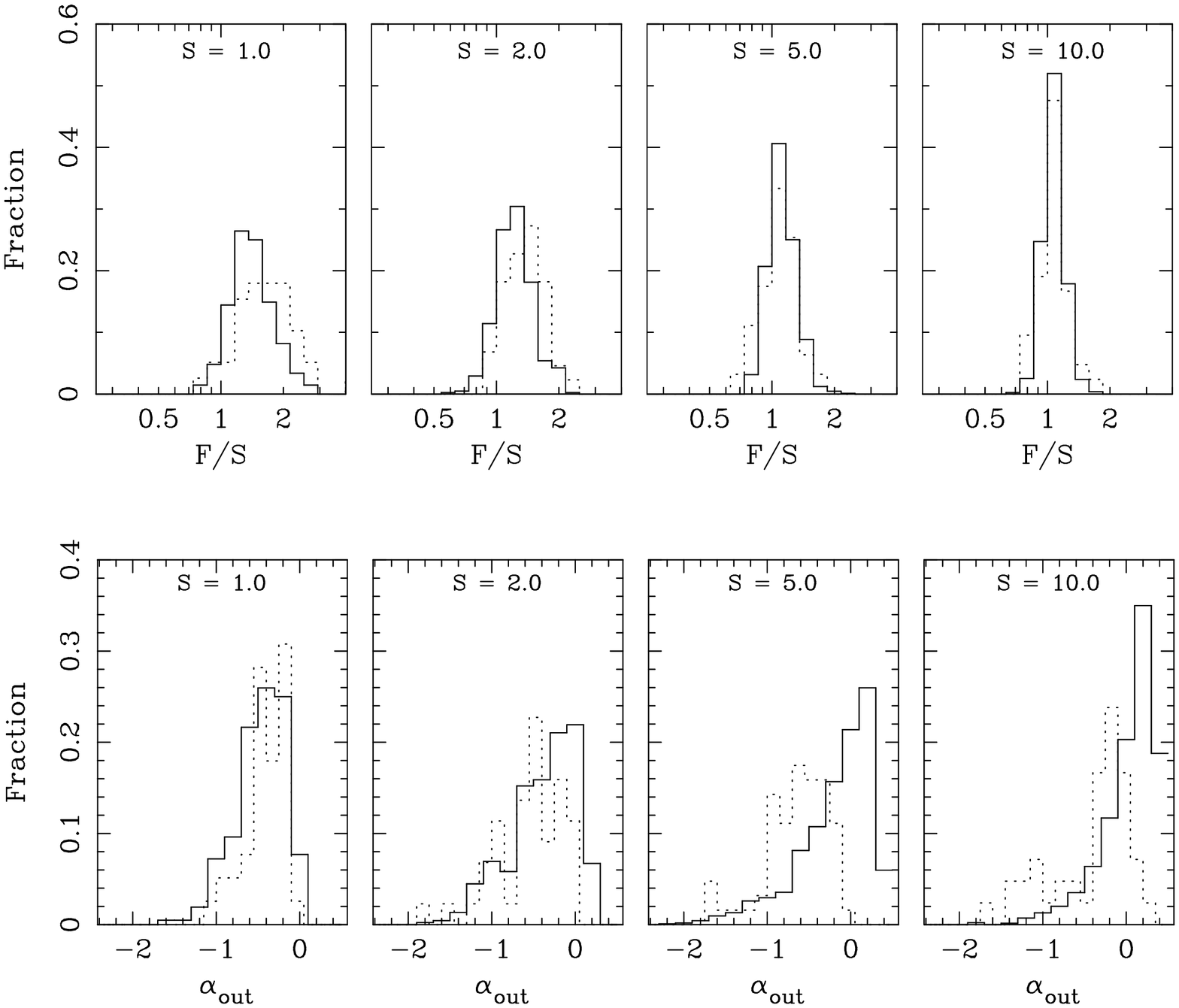,width=170truemm,angle=270}
\caption{Distributions of $F/S$ (top panel, where $S$ is the intrinsic flux and
$F$ is the fitted flux) and fitted spectral slope (bottom panel) for simulated
marginally hard sources ($\alpha = 0.5$, solid histograms) and soft sources
($\alpha = 1$, dashed histograms) which achieve the spectral selection
criterion (Eq. \ref{eq:spectralcriterion}). Note that each histogram is
normalised, and does not represent the sensitivity of the survey to those
sources, which is given by Fig. \ref{fig:effareas}.}
\label{fig:twosoft}
\end{center}
\end{figure*}

Fig. \ref{fig:twosoft} shows the distributions of \(F/S\) and fitted spectra
for marginally hard sources (\(\alpha=0.5\), solid histograms) and ordinary
sources (\(\alpha=1\), dashed histogram) scattered into the hard survey by
poisson noise (see Section \ref{sec:contaminants}). The simulated faint
\(\alpha=1\) sources which are scattered into the survey show a particularly
strong skew towards larger \(F/S\); this is unfortunate because it increases
the expected contamination of the sample by 'normal' sources (Section
\ref{sec:contaminants}).  At bright fluxes the \(\alpha=0.5\) sources enter the
survey with a softer distribution of slopes than either the hard sources
(\(\alpha \leq 0\)) or the scattered \(\alpha=1\) sources.  At faint fluxes
(\(S \sim 10^{-14}\) \ecs) the \(\alpha=0.5\) sources and scattered
\(\alpha=1\) sources have distributions of fitted slope which are similar to
each other and to those of harder sources.  All these effects will be taken 
into account in Section \ref{sec:spectrum} when we characterise the spectra of 
the real hard sources. 

\section{Results}
\label{sec:results}

The catalogue of 147 serendipitous \ros\ hard sources which comprise our hard
source sample is presented in Table \ref{tab:hardcat}. Source names are based
on the \ros\ positions derived from PSS and are given in column 1. Column 2
gives the name of the \ros\ observation in which the hard source was detected,
and column 3 gives the PSPC exposure time after screening (Section
\ref{sec:reduction}). The offaxis angle of the source is given in column 4. The
positional uncertainty of the X-ray source, as determined by PSS, is given in
column 5, and the PSS significance of the source (number of sigmas above the
background level) is given in column 6.  As an indicator of whether or not the
source is point-like, \(\chi^{2}/\nu\) (where \(\nu=11\)) from fitting the
source radial profile with the PSPC point spread function (Section
\ref{sec:radial}) is given in column 7.  The size of the source circle used to
extract the 3 colour spectrum (Section \ref{sec:reduction}) is given in column
8. The numbers of counts within this radius for each of the 3 bands (channels
11-41, 52-90 and 91-201) are given in columns 9, 10 and 11 respectively, and
the the predicted numbers of background counts for the same bands are given in
columns 12, 13 and 14.  The fitted spectral slope for the source and 68\%
uncertainty (Section \ref{sec:fitting}) are given in column 15. The probability
that the source has a spectrum harder than \(\alpha=0.5\) (Section
\ref{sec:criteria}) is given in column 16, and the fitted 0.5 - 2.0 keV
flux of the source, with 68\% errors (Section \ref{sec:fitting}) is given in
column 17. 
Finally, column 18 contains a flag as to whether the source has a likely
optical counterpart and will be used in subsequent papers about optical
spectroscopy and imaging of the hard sources.  `S' means that the source
is suitable for optical spectroscopy, while `I'
means that we consider the source only suitable for imaging follow up.

This sample represents the detection of a significant
population of hard sources in \ros. Before we investigate their \(N(S)\) 
relation, we will demonstrate that the contamination of the sample from 
non-hard sources is low, and determine an approximate characteristic 
spectral slope for the the hard source sample.

\subsection{Contamination of the survey by non-hard sources}
\label{sec:contaminants}

\begin{figure}
\begin{center}
\leavevmode
\psfig{figure=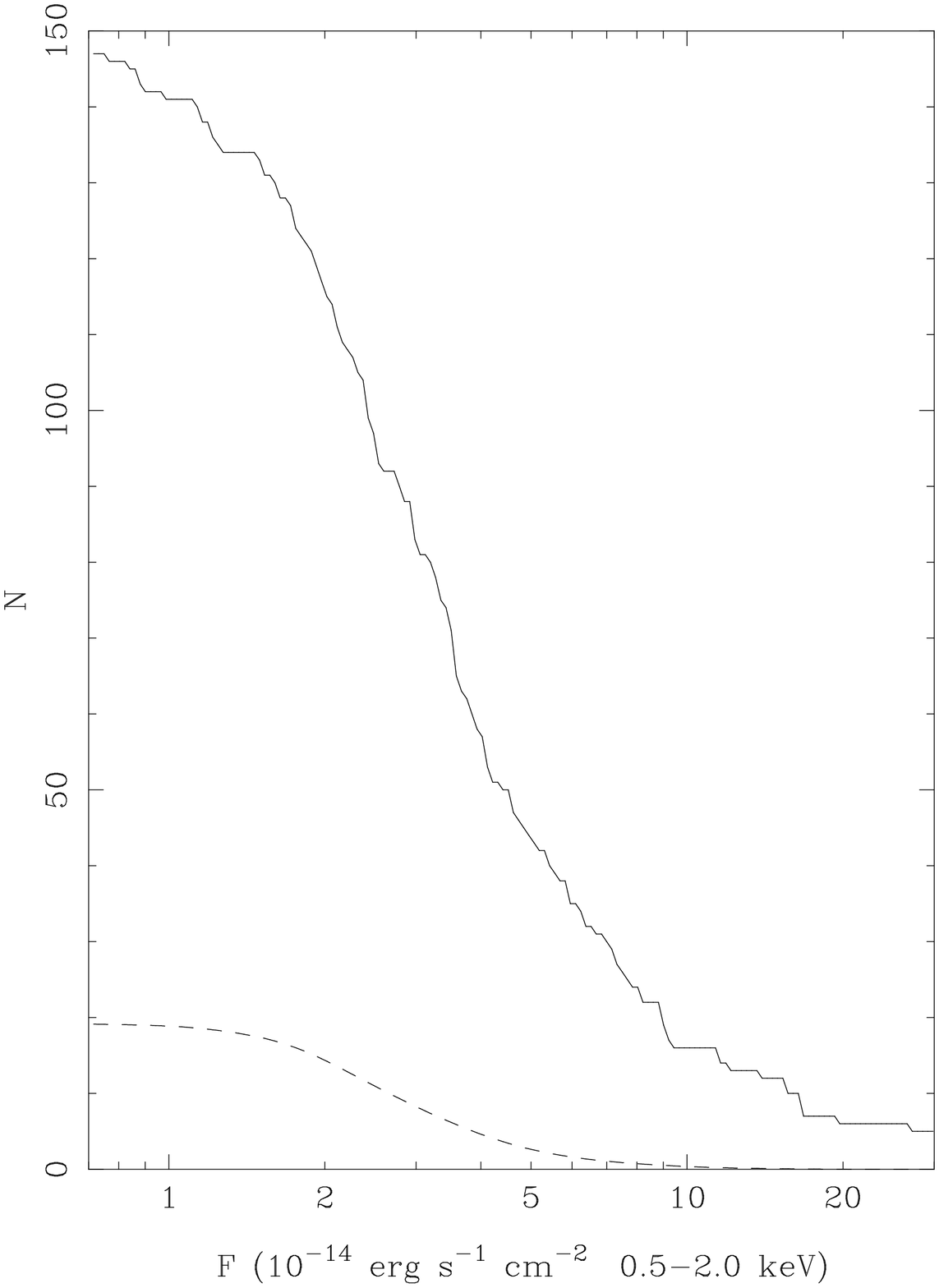,width=80truemm}
\caption{Worst case predicted number of contaminant $\alpha \sim 1$ sources
(dashed line, see Section \ref{sec:contaminants}) compared to the actual
content of the survey (solid line) as a function of observed flux $F$.}
\label{fig:contaminants}
\end{center}
\end{figure}

Although Fig. \ref{fig:effareas} shows that the survey is much more efficient
at detecting hard sources than soft sources, the source population at the flux
levels probed by \ros\ is dominated by sources which are softer than the
background, with a mean spectrum of \(\alpha \sim 1\) (\eg Branduardi-Raymont
\etal 1994, Hasinger \etal 1993b). Inevitably, poisson noise will scatter some
soft sources into the hard survey; this is why the effective area of the survey
to \(\alpha=1.0\) sources is non-zero.  We constructed a `worst case' estimate
for the number of ordinary (\(\alpha \sim 1\)) sources scattered into the hard
source survey by multiplying the \(\alpha = 1\) effective area curve in
Fig. \ref{fig:effareas} by the \(N(S)\) relation of the entire X--ray
source population (from Branduardi-Raymont et al 1994) and convolving with the
\(\alpha = 1\) distribution of \(F/S\) (see Section
\ref{sec:recovery}). The resultant predicted number of contaminant sources, as
a function of flux, is compared to the actual source counts in
Fig. \ref{fig:contaminants}. The predicted number of soft sources scattered
into the survey reaches 19 (13\% of the total) at the faint limit of our survey
(\(\sim 10^{-14}\) \ecs).

In reality we expect the number of `normal' soft sources scattered into our
survey to be considerably smaller than this because: 

\noindent 1) The total \(N (> S)\) includes hard sources which are not a
contaminant.

\noindent 2) The majority of the sources at \(S > 10^{-14}\) \ecs\ 
are AGN, and the more than half of these are actually softer than
\(\alpha = 1\) (eg see figure 7 of Mittaz \etal 1999) and so will be even more
efficiently rejected by the spectral selection than the simulated \(\alpha =
1\) sources.

We therefore expect that the level of contamination of the hard source sample 
by normal soft sources is small, \(<13\%\).

\subsection{Characteristic X-ray spectral slope of the hard sources}
\label{sec:spectrum}

The hard source survey has significant effective area to sources with almost
any spectral slope \(\alpha \leq 0.5\), but just {\it how hard} the source
spectra are has a considerable bearing on their contribution to the XRB, and
the possible origins of their X-ray spectra.  Many of the sources are faint,
with fairly large uncertainties on the fitted \(\alpha\), which will be biased
by the spectral selection criterion (see Section \ref{sec:recovery}).  The
sample probably contains objects with a range of spectra, but in this section
we characterise our hard sample with a representative spectral
slope. We determine this by looking for a value of intrinsic slope \(\alpha\)
which would give rise to the observed distribution of fitted spectral slopes in
the hard source sample.

Once again, we use our Monte Carlo simulations described in Section
\ref{sec:recovery}.  For each value of the input spectral slope \(\alpha\), a
distribution of fitted spectral slopes appropriate for the flux \(F\) of each
hard source in the sample, was constructed from the simulations.  This was done
by linearly interpolating between the distributions of fitted slopes at
discrete simulated fluxes.  A simulated distribution of fitted spectra
appropriate for the sample as a whole (for each value of intrinsic \(\alpha\))
was then obtained by summing the simulated distributions for the individual
hard sources.

\begin{figure}
\begin{center}
\leavevmode
\psfig{figure=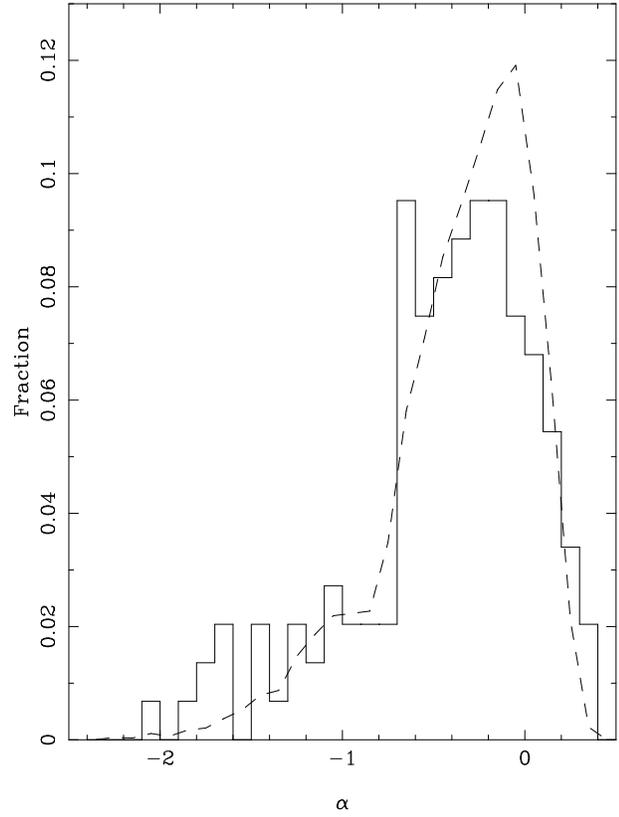,width=80truemm}
\caption{Distribution of hard source spectral slopes (solid histogram) and the
 simulated
distribution for $\alpha=0$ sources (dashed line).}
\label{fig:realspecdist}
\end{center}
\end{figure}

To compare these simulated distributions of fitted
spectral slopes with the real distribution, we used the Kolmogorov Smirnov 
(KS) test. The results of these comparisons are given in Table 
\ref{tab:ksresults}. 
The spectral slope which reproduces the distribution
of fitted slopes best is \(\alpha=0\); all the other values for
\(\alpha\) are rejected by the KS test with \(>99.99\%\) confidence.
Binned versions of the real distribution of spectral
slopes and the intrinsic \(\alpha=0\) simulated distribution are shown in
Fig. \ref{fig:realspecdist}. 
Our hard source sample as a whole is therefore best
characterised by a spectral slope of \(\alpha \sim 0\); the sample is neither
dominated by sources with similar spectra to that of the XRB (\(\alpha \sim
0.5\)) nor by sources which are \(\it extremely\) hard (\(\alpha \leq -0.5\)).

\begin{table}
\caption{KS test results comparing the distribution of fitted spectral slopes
in the real sample, against distributions simulated assuming different
intrinsic spectral slopes $\alpha$}
\label{tab:ksresults}
\begin{tabular}{ccc}
$\alpha$& D$_{\rm max}$ & P $>$ D$_{\rm max}$\\
-1.0&0.40&6.9$\times 10^{-21}$\\
-0.5&0.22&7.2$\times 10^{-7}$\\
0.0&0.10&0.12\\
0.5&0.20&1.5$\times 10^{-5}$\\
\end{tabular}
\end{table}

\subsection{Source counts}
\label{sec:sourcecounts}

The hard sources \(N(S)\) (sky density of sources per unit flux, at flux S) is
the critical issue as to whether the population of \ros\ hard sources are
likely to contribute significantly to the XRB, and the solution to 
the spectral paradox,
at faint fluxes. To derive the hard source \(N(S)\), we use the \(\alpha=0\)
effective area curve from Fig. \ref{fig:effareas}, since we have already shown
that \(\alpha=0\) best typifies the source spectra 
(Section \ref{sec:spectrum}).
A power law model fit to the \(N(S)\) relation was obtained using the Murdoch,
Crawford and Jauncey (1973) maximum likelihood method for sources with
measurement uncertainty. This is appropriate because we know from our
simulations that the fitted fluxes (\(F\)) of the sources deviate from the 
actual fluxes (\(S\))
by more than 20\% (see the top panel of Fig. \ref{fig:twohard}). The
method works by convolving the model \(N(S)\) with the error distribution of
the sample, to produce a model probability distribution of observed fluxes
\(P(F)\); the fitting proceeds by adjusting the shape of \(N(S)\), hence
\(P(F)\), to maximise the likelihood of obtaining the sample's observed flux
distribution.

The power law model \(N(S)\) is defined as:
\[
N(S) = K\,S^{-\gamma}
\]
within the interval \(S_{min}<S<S_{max}\).
This is transformed to the model probability density
\(P(F)\) that a source in the sample will have observed flux F by:
\[
P(F) = \frac
{\int_{S_{min}}^{S_{max}}P(F \mid S)\ N(S)\ A(S)\ dS}
{\int_{F_{min}}^{F_{max}}
\int_{S_{min}}^{S_{max}}P(F \mid S)\ N(S)\ A(S)\ dS\ dF}
\] 
where
\(P(F \mid S)\) is the probability density of observed flux \(F\) given
intrinsic flux \(S\) and \(A(S)\) is the effective area to sources of intrinsic
flux \(S\). \(P(F \mid S)\) and \(A(S)\) were obtained from the Monte Carlo
simulations. \(F_{min}<F<F_{max}\) is the interval of observed
flux in which the fitting was performed.  To ensure that \(P(F)\) is correctly
determined close to \(F_{min}\) and \(F_{max}\), \(S_{min}\) and \(S_{max}\)
were set to 50\% and 200\% of \(F_{min}\) and \(F_{max}\) respectively.  We
chose to fit the \(N(S)\) with \(F_{min} = 10^{-14}\) \ecs, 
to minimise any contribution from sources fainter than the minimum
flux of the effective area simulations, and \(F_{max} = 2 \times 10^{-13}\) 
\ecs, above which the sample may be incomplete due to our exclusion of
\ros\ observation targets.
The fit was performed by varying \(\gamma\) to maximise
\[
P = \prod_{i=1}^{n} P(F_{i})
\]
where n is the number of sources with flux \(F_{i}\) with
\(F_{min}<F_{i}<F_{max}\).
Estimates for the 68\% and 95\% uncertainty of \(\gamma\) were obtained by
finding values for \(\gamma\) corresponding to  
\(\Delta (-2\,{\rm log}(P))\) = 1 and 4 respectively.
The normalisation \(K\) was determined by setting the number of observed
sources to the number predicted:
\[
K = \frac
{n}
{\int_{F_min}^{F_max}\int_{S_min}^{S_max}P(F \mid S)\ S^{-\gamma}\ A(S)\ dS\
dF}
\]

\begin{figure}
\begin{center}
\leavevmode
\psfig{figure=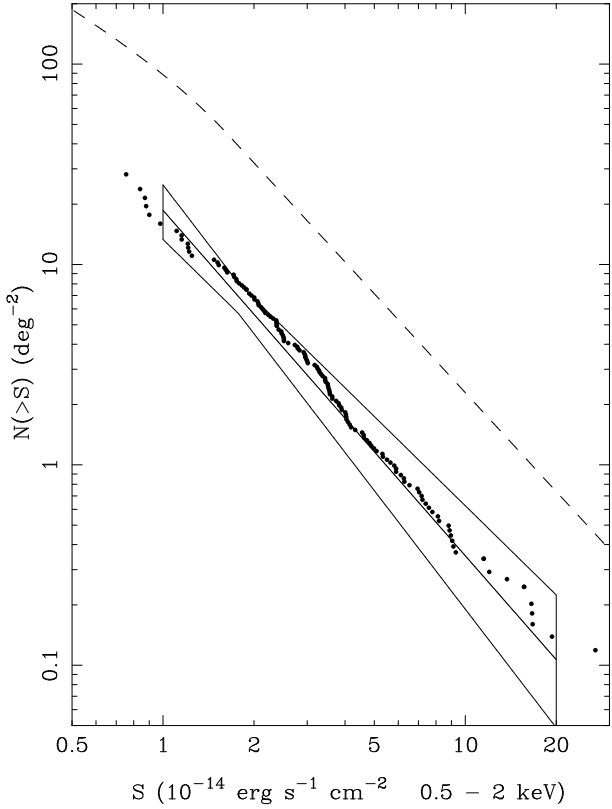,width=80truemm}
\caption{Best fit integral $N(>S)$ for the hard sources (solid line)
along with 95\% confidence interval (bowtie) and crude $N(>S)$ obtained
by summing the inverse of the area available to each source (dots). Also shown
for comparison is the $N(>S)$ of the whole population (dashed line,
from Branduardi-Raymont \etal 1994)}
\label{fig:hardns}
\end{center}
\end{figure}

The best fit value of \(\gamma\) is \(2.72 \pm 0.12\) (\(\pm 0.24\) at 95\%)
with a normalisation \(K=32\) (\(10^{-14}\) \ecs)\(^{(\gamma
-1)}{\rm deg^{-2}}\).  The best fit model is shown in Fig. \ref{fig:hardns}
along with the uncertainty bowtie corresponding to the 95\% maximum likelihood
fitting errors as well as a 95\% normalisation error based on the number of
sources in the sample.
For reference, Fig. \ref{fig:hardns} also shows a crude \(N(>S)\) for the hard
sources obtained by 
\[
N(>S) \sim \sum^{n}_{i=1} 1/A(F_{i})\ \ (F_{i}>S)
\] (dots)
as well as the best fit model \(N(>S)\) for the overall source population 
(dashed line, from Branduardi-Raymont \etal 1994).

The best fit hard source \(N(S)\) slope is steeper than the Euclidean value of
\(\gamma=2.5\) and the model normalisation translates to \(N (> 10^{-14}\)
\ecs\() = 19 \) deg\(^{-2}\).  The hard sources are therefore a
significant component (\(>10\%\)) of the \(S > 10^{-14}\) \ecs\ 0.5 - 2 keV
population. 

\subsection{Contribution to the 1-2 keV XRB}
\label{sec:xrb}

We have estimated the hard sources contribution to the 1-2 keV XRB, \(I_{1-2}\)
from the
fitted power law \(N(S)\) from Section \ref{sec:sourcecounts} assuming a
spectral slope of \(\alpha=0\) (see Section \ref{sec:spectrum}).  
\[
I_{1-2}=\frac{2K(S_{min}^{(2-\gamma)}-S_{max}^{(2-\gamma)})}{3(2-\gamma)}
\]
where \(K\) and \(\gamma\) are the normalisation and slope of the \(N(S)\) as
defined in Section \ref{sec:sourcecounts} and \(S_{min}\) and \(S_{max}\)
denote the 0.5 - 2 keV flux range of hard sources considered. We have assumed 
\(S_{max} = 10^{-11}\) \ecs.

For \(S_{min} =
10^{-14}\) \ecs\ (0.5 - 2 keV) the hard sources are responsible for a 1 - 2 keV
intensity of \(I_{1-2}=1.0^{+0.3}_{-0.2} \times 10^{-9}\) erg s\(^{-1}\)
cm\(^{-2}\) sr\(^{-1}\) (95\% errors).
Assuming the 1 - 2 keV XRB intensity is \(\sim 1.45 \times 10^{-8}\) erg
s\(^{-1}\) cm\(^{-2}\) sr\(^{-1}\), from the joint \asca\ \ros\ spectral fits
of Chen, Fabian \& Gendreau (1997) and Mijayi \etal (1998), the hard sources 
contribute \(7\pm 2 \%\) of
the 1 - 2 keV background.
Extrapolating the hard source \(N(S)\) relation to \(S = 10^{-15}\) \ecs\ (0.5
- 2 keV), they would produce an intensity of \(I_{1-2}=5.1^{+5.1}_{-2.5}
\times 10^{-9}\) erg s\(^{-1}\) cm\(^{-2}\) sr\(^{-1}\), or
\(35^{+36}_{-16}\%\) of the background.  The hard sources therefore have the 
potential to be major contributors to the XRB at faint fluxes.

\section{Discussion}
\label{sec:discussion}

With the construction of our hard source catalogue, we have isolated a
population of sources which have hard spectra (\(\alpha \sim 0\)), have a steep
\(N(S)\), and are numerous enough to make up \(\sim 15\%\) of
sources with \(S > 10^{-14}\) \ecs.  The total \(N(S)\) of all sources in the
0.5 - 2 keV band has already flattened off to a sub-Euclidean slope by
\(10^{-14}\) \ecs\ (Branduardi-Raymont \etal 1994, Hasinger \etal 1993).
Extrapolating the hard source \(N(S)\) shown in
Fig. \ref{fig:hardns} to fainter fluxes, the hard source contribution increases
to around 40\% of all the sources between 0.5 and 1 \(\times 10^{-14}\)
\ecs. This provides a very simple explanation for the hardening of the mean
\ros\ source spectrum towards faint fluxes found by Mittaz \etal (1999), 
Vikhlinin \etal (1995) and Hasinger \etal (1993b).

The results presented here also offer a consistent picture of the source
populations found in the \ros\ band and at higher energy.  A long standing
discrepancy has been the excess source counts in the 2-10 keV band compared to
those in the 0.5-2 keV band (Warwick \& Stewart 1989, Butcher \etal 1997).
From the \asca\ Large Sky Survey and \asca\ Medium-Sensitivity Survey, Ueda
\etal (1999a) and (1999b) recently showed that in the 0.7-7 keV energy range,
sources with spectra harder than \(\alpha=0.7\) have a steeper \(N(S)\) 
than softer spectrum sources. These harder sources make a small contribution to
the source counts at very bright fluxes (\(<20\%\) at \(> 10^{-12}\) \ecs) 
but because of their steep \(N(S)\) are
almost as numerous as the soft sources at \(S < 10^{-13}\) \ecs\ 
(0.7-7 keV).  Although the hard sources found in the 2-10 keV band
have average energy spectra of \(\alpha \sim 0.5 - 0.6\) (Ueda \etal 1999a,
1999b), softer than the \ros\ hard source spectra, they play a similar role in
hardening the average source spectrum at faint fluxes. 

The \asca\ hard sources become significant at much brighter fluxes than the
\ros\ hard sources. This is to be expected because hard spectrum sources
contribute to higher energy source counts at brighter fluxes than soft sources;
\eg a source with \(\alpha=0\) is more than four times brighter in the 2-10 keV
band than an \(\alpha=1\) source with the same 0.5 - 2 keV flux.  Extrapolating
the \ros\ hard source counts assuming \(\alpha=0\) results in \(\sim 5\)
sources deg\(^{-2}\) at \(10^{-13}\) \ecs (2 - 10 keV).  The sky density of 2 -
10 keV sources at this flux level is between 10 and 20 deg\(^{-2}\) (Ueda \etal
1999a, 1999b, Georgantopoulos \etal 1998, Inoue \etal 1996).  This means that
the hard \ros\ population found by our survey could be a significant
contributor to the 2 - 10 keV source counts, and that the \ros\ hard source
population could constitute a large fraction of the \asca\ hard sources as
well.  This hypothesis is supported by Ueda \etal (1999a and 1999b) finding
that \(\sim 80\%\) of the \asca\ sources detected in 2 - 10 keV are also
detected in 0.7 - 2 keV, which implies that the population of hard \asca\
sources {\it should} be found in the \ros\ band as well.

\section{Conclusions}
\label{sec:conclusions}

We have performed a survey of 188 \ros\ fields for sources with hard spectra
(\(\alpha < 0.5\)), and present a serendipitous catalogue of 147 such hard
sources. We have applied our spectral selection criterion to Monte Carlo
simulations to calculate the effective area of our survey as a function of
source flux and spectral slope. The Monte Carlo simulations have also been used
to estimate biases in fitted flux and spectral slope resulting from the hard
spectral selection.  The effective area of the survey is at least 10 times
greater for hard sources than for `normal' \(\alpha \sim 1\) sources for 0.5 -
2 keV source flux \(S > 10^{-14}\) \ecs.  Convolving the overall \(N(S)\)
of 0.5 - 2 keV sources with the selection function of our survey, we show
that the level of contamination of the sample by normal \(\alpha \sim 1\)
sources is small (\(\leq 15\%\)).  The distribution of hard sources fitted
spectral slopes implies that a typical source in our sample has a spectral
slope \(\alpha \sim 0\).  The hard sources have a steep \(N(S)\) relation
(\(dN/dS \propto S^{-\gamma}\) with a best fit value of \(\gamma = 2.72\pm
0.12\)) and make up about 15\% of all 0.5 - 2 keV sources with \(S > 10^{-14}\)
\ecs. If their \(N(S)\) continues to fainter fluxes, the hard
sources will comprise \(\sim 40\%\) of sources with \(5 \times 10^{-15}\) \ecs
\(< S <\) \(10^{-14}\) \ecs.  The increased contribution of hard sources to the
faint \ros\ population can therefore account for the harder average spectra of
\ros\ sources with \(S < 10^{-14}\) \ecs. The \ros\ hard source sample is
probably the bright tail of the population that contributes much of the soft
(and perhaps hard) X-ray background at faint fluxes. The \ros\ hard sources are
probably a subset of the hard source population now detected in higher energy
\asca\ observations.

\section{Acknowledgments}

FJC thanks the DGES for partial financial support, under project PB95-0122.
This research has made use of data obtained from the Leicester Database and
Archive Service at the Department of Physics and Astronomy, Leicester
University, UK.  This research has also made use of the NASA/IPAC Extragalactic
Database (NED) which is operated by the Jet Propulsion Laboratory, California
Institute of Technology, under contract with the National Aeronautics and Space
Administration. 
We thank Dr Andrew Phillips for constructing the MSSL `Beowulf'
parallel computer which was invaluable for our Monte Carlo simulations.

\section{References}

\refer Almaini O., Shanks T., Boyle B.J., Griffiths R.E., Roche N., Stewart
G.C., Georgantopoulos I., 1996, MNRAS, 282, 295

\refer Almaini O., Shanks T., Griffiths R.E., Boyle B.J., Roche N.,
Georgantopoulos I., Stewart G.C., 1997, MNRAS, 291, 372

\refer Branduardi-Raymont G., \etal, 1994, MNRAS, 270, 947

\refer Butcher J.A., \etal 1997, MNRAS, 291, 437

\refer Chen L.W., Fabian A.C., Gendreau K.C., 1997, MNRAS, 285, 449

\refer Ciliegi P., Elvis M., Wilkes B.J., Boyle B.J., McMahon R.G., Maccacaro
T., 1997, MNRAS, 284, 401

\refer Georgantopoulos I., \etal 1997, MNRAS, 291, 203

\refer Giacconi R. \etal, 1962, Phys. Rev. Letters, 9, 439.

\refer Hasinger G., Boese G., Predehl P., Turner T.J., Yusaf R., George I.M.,
Rohrbach G., 1993a, MPE/OGIP Calibration Memo CAL/ROS/93-015

\refer Hasinger G., Burg R., Giacconi R., Hartner G., Schmidt M., Tr\" umper
J., Zamorani G., 1993b, A\&A, 275, 1

\refer Hasinger G., Burg R., Giacconi R., Schmidt M., Tr\" umper J., Zamorani
G., 1998, A\&A, 329, 482

\refer Inoue H., Kii, T., Ogasaka Y., Takahashi T., Ueda Y., 1996, 
in Zimmerman U.,  Tru\" mper J.E.,  Yorke H., eds, Ro\" ntgenstrahlung from the
Universe, MPE Report 263, p. 323

\refer Mason K.O. \etal, 2000, MNRAS, in press

\refer Miyaji T., Ishisaki Y., Ogasaka Y., Ueda, Y., Freyberg M.J., Hasinger
G., Tanaka Y., 1998, A\&A, 334, L13

\refer Mittaz J.P.D. \etal, 1999, MNRAS, 308, 233

\refer Newsam, A.M., McHardy I.M., Jones L.R., Mason K.O., 1999, MNRAS, 310,
255

\refer Roche N., Griffiths R.E., della Ceca R., Shanks T., Boyle B.J., 
Georgantopoulos I., Stewart G.C., 1996, MNRAS, 282, 820

\refer Romero Colmenero E., Branduardi-Raymont G., Carrera F.J., Jones L.R.,
Mason K.O., McHardy I.M., Mittaz J.P.D., 1996, MNRAS, 282, 94

\refer Schmidt M., Hasinger G., Gunn J., Schneider D., Burg R., Giacconi R.,
Lehmann I., MacKenty J., Tr\" umper J., Zamorani G., 1998, A\&A, 329, 495

\refer Snowden S.L., McCammon D., Burrows D.N., Mendenhall J.A., 1994, ApJ,
424, 714

\refer Vikhlinin A., Forman W., Jones C., Murray S., 1995, ApJ, 451, 564

\refer Warwick R.S., Stewart G.C., 1989, ESA SP-296, 2, 727

 \begin{table*}
 \tabcolsep=1mm
 \caption{Hard source catalogue}
 \label{tab:hardcat}
 \begin{tabular}{llllllllllllllcccc}
 1&2&3&4&5&6&7&8&9&10&11&12&13&14&15&16&17&18\\
                                 Source      & ROR & Exp &Offax & Perr & $\sigma$ & PSF & r & c1 & c2 & c3 & b1 & b2 & b3 & $\alpha$ & P& Flux & Spl\\
 &&(ks)&$(')$&$('')$&&$\chi^{2}_{/\nu}$&$('')$&&&&&&&&$<0.5$&0.5-2&\\
 &&&&&&&&&&&&&&&&&\\
RXJ000417.24-255915.1&rp700467n00&31.7&12.7&11.2& 5.2&  1.6&54&  51&  26&  32& 66.4&  9.9&  7.9&-0.45$^{+0.66}_{-0.56}$&0.99&  1.6$^{ +0.4}_{ -0.3}$&I\\
RXJ000721.30+204326.5&rp700101n00&20.8&13.7& 6.6& 4.9&  1.2&43&  20&   9&  20& 20.5&  4.5&  3.4&-0.60$^{+1.06}_{-1.60}$&0.92&  1.8$^{ +0.6}_{ -0.5}$&I\\
RXJ001111.73-361840.1&rp800388n00&20.9& 6.7& 6.2& 8.8&  0.5&54&  48&  14&  29& 40.7&  5.1&  3.9&-0.28$^{+0.68}_{-0.75}$&0.99&  2.2$^{ +0.5}_{ -0.4}$&I\\
RXJ001144.43-362638.0&rp800388n00&20.9& 5.3& 3.4&15.6&  1.1&54&  54&  27&  50& 40.7&  5.1&  3.9&-0.14$^{+0.27}_{-0.35}$&1.00&  4.2$^{ +0.6}_{ -0.6}$&S\\
RXJ004651.98-204329.0&rp600159a00& 9.0& 4.6& 1.6&31.3&  2.1&54&  59&  74&  79& 16.4&  3.0&  2.3& 0.33$^{+0.13}_{-0.11}$&0.94& 19.4$^{ +2.0}_{ -1.6}$&S\\
RXJ005726.44+270124.4&rp700884n00&18.9&15.8& 7.3& 9.4&  0.9&54&  19&  15&  45& 16.7&  6.7&  8.2&-0.91$^{+1.10}_{-1.02}$&0.97&  4.6$^{ +1.1}_{ -0.8}$&I\\
RXJ005732.46-273005.8&rp701223n00&43.7& 8.3& 6.9& 6.0&  1.7&25&  18&   6&  19& 16.2&  2.4&  1.7&-0.90$^{+1.22}_{-1.38}$&0.98&  0.8$^{ +0.2}_{ -0.2}$&I\\
RXJ005734.78-272827.4&rp701223n00&43.7&10.0& 3.9& 9.1&  1.4&50&  59&  19&  32& 57.5&  8.5&  5.9&-0.34$^{+0.56}_{-0.99}$&0.97&  1.2$^{ +0.3}_{ -0.2}$&S\\
RXJ005736.38+302355.2&rp700424a01&17.0& 3.8& 5.7& 8.0&  1.7&40&   5&   7&  19&  9.4&  2.8&  2.0&-0.89$^{+0.92}_{-1.61}$&0.96&  1.9$^{ +0.6}_{ -0.5}$&S\\
RXJ005736.81-273305.9&rp701223n00&43.7& 5.6& 6.8& 7.1&  0.5&50&  62&  26&  33& 60.6&  8.9&  6.2& 0.00$^{+0.36}_{-0.64}$&0.94&  1.2$^{ +0.2}_{ -0.2}$&S\\
RXJ005746.75-273000.8&rp701223n00&43.7& 9.3& 4.1& 8.7&  1.5&54&  62&  39&  41& 63.0&  9.3&  6.4& 0.08$^{+0.34}_{-0.32}$&0.95&  1.7$^{ +0.3}_{ -0.3}$&S\\
RXJ005801.64-275308.6&rp701223n00&43.7&16.4& 5.2&14.7&  1.0&54&  67&  47&  99& 57.4&  8.4&  5.9&-0.36$^{+0.31}_{-0.29}$&1.00&  5.1$^{ +0.5}_{ -0.5}$&S\\
RXJ005808.69-273535.2&rp701223n00&43.7& 9.3& 5.0& 6.1&  1.1&36&  23&  13&  21& 26.5&  3.9&  2.7&-0.36$^{+0.71}_{-0.74}$&0.97&  0.9$^{ +0.2}_{ -0.2}$&I\\
RXJ005812.20-274217.8&rp701223n00&43.7&10.4& 7.7& 5.9&  1.0&54&  58&  20&  26& 63.2&  9.3&  6.5&-0.30$^{+0.58}_{-1.02}$&0.96&  0.9$^{ +0.2}_{ -0.2}$&I\\
RXJ005830.02+263921.9&rp700884n00&18.9&12.9& 2.3&21.5&  1.6&54&  21&  25& 116& 18.0&  7.3&  8.8&-1.69$^{+0.70}_{-0.56}$&1.00& 11.5$^{ +1.6}_{ -1.2}$&S\\
RXJ010742.15+322642.5&rp600106n00&23.4& 4.4& 3.0&11.6&  3.6&18&   5&  11&  38&  4.9&  2.4&  3.4&-0.64$^{+0.82}_{-0.84}$&0.96&  3.5$^{ +0.8}_{ -0.5}$&S\\
RXJ010908.73+192759.5&rp201488n00&12.9&14.8&12.2& 5.3&  1.7&50&   9&   8&  17& 13.1&  3.1&  2.7&-0.56$^{+0.98}_{-1.30}$&0.94&  2.5$^{ +0.9}_{ -0.6}$&S\\
RXJ013555.47-183210.2&rp200208n00&24.8&15.1& 8.5& 8.8&  0.9&54&  55&  21&  41& 49.0&  8.1&  4.5&-0.53$^{+0.45}_{-0.75}$&1.00&  3.2$^{ +0.6}_{ -0.6}$&S\\
RXJ013707.63-183846.4&rp200208n00&24.8&17.6&10.0&10.2&  1.1&54&  55&  26&  44& 45.3&  7.5&  4.1&-0.23$^{+0.30}_{-0.47}$&1.00&  4.0$^{ +0.7}_{ -0.6}$&S\\
RXJ013717.52-182716.9&rp200208n00&24.8& 9.6& 4.0& 8.6&  1.0&18&   8&   4&  19&  5.9&  1.0&  0.5&-0.98$^{+1.12}_{-1.33}$&1.00&  2.5$^{ +0.7}_{ -0.6}$&I\\
RXJ013721.47-182558.3&rp200208n00&24.8& 9.8& 2.6&15.4&  1.6&36&  37&  31&  46& 22.6&  3.8&  2.1& 0.03$^{+0.19}_{-0.27}$&0.99&  3.9$^{ +0.6}_{ -0.5}$&S\\
RXJ014159.22-543037.0&rp800383n00& 7.4&11.8& 3.4&11.9&  0.9&54&  21&  18&  36& 18.4&  3.1&  3.7&-0.22$^{+0.41}_{-0.57}$&0.98&  8.8$^{ +1.5}_{ -1.5}$&S\\
RXJ021658.11-173203.3&rp900352n00& 9.1&14.4&14.8& 5.3&  0.8&54&   7&   9&  15&  9.8&  1.5&  1.5&-0.35$^{+0.51}_{-0.94}$&0.96&  3.3$^{ +0.8}_{ -0.9}$&I\\
RXJ023313.71+005536.0&rp800482n00&24.9&12.6& 6.2& 9.3&  3.1&54&  36&  19&  34& 35.9&  4.4&  4.2&-0.06$^{+0.45}_{-0.65}$&0.93&  2.5$^{ +0.4}_{ -0.4}$&S\\
RXJ023851.91-520958.8&rp701356n00&21.3&16.3&13.1& 7.6&  1.5&54&  36&  24&  48& 39.2&  5.0&  5.5&-0.25$^{+0.50}_{-0.50}$&0.98&  4.9$^{ +0.9}_{ -0.6}$&S\\
RXJ025753.52+194442.2&rp900138n00&21.2&12.8& 3.3&17.1&  1.1&54&  30&  27&  92& 28.7&  4.0&  2.9&-0.53$^{+0.50}_{-0.57}$&0.98&  9.3$^{ +1.1}_{ -0.9}$&I\\
RXJ025755.08+194255.5&rp900138n00&21.2&14.0& 4.7&14.7&  0.8&54&  33&  15&  88& 28.0&  3.9&  2.8&-1.86$^{+0.75}_{-0.79}$&1.00&  9.1$^{ +1.4}_{ -1.0}$&I\\
RXJ031043.21-472033.4&rp800304n00& 7.7& 8.7&14.5& 4.3&  1.4&54&  17&   6&  12& 18.3&  2.2&  2.3&-0.54$^{+0.89}_{-1.50}$&0.95&  2.4$^{ +0.9}_{ -0.8}$&S\\
RXJ031114.91-473150.0&rp800304n00& 7.7& 3.9& 9.8& 5.6&  0.7&47&  13&   5&  10& 13.2&  1.6&  1.7&-0.41$^{+0.89}_{-1.48}$&0.93&  2.0$^{ +0.9}_{ -0.7}$&I\\
RXJ031456.58-552006.8&rp701036n00&40.6& 6.6& 3.6&12.6&  0.9&54& 101&  23&  82& 93.3&  9.6&  6.1&-1.66$^{+0.70}_{-0.85}$&1.00&  3.5$^{ +0.5}_{ -0.5}$&S\\
RXJ031956.76-663938.5&rp600504n00&14.1&14.0&60.0&15.4&  1.6&54&  16&  23&  64& 12.9&  3.5&  3.9&-0.44$^{+0.50}_{-0.48}$&1.00&  8.9$^{ +1.3}_{ -1.0}$&I\\
RXJ033340.22-391833.4&rp800367a01&14.8&14.5& 7.6& 8.5&  1.4&54&  37&  12&  29& 30.0&  3.7&  3.7&-0.24$^{+0.71}_{-0.77}$&0.98&  3.4$^{ +0.7}_{ -0.7}$&S\\
RXJ033350.25-390850.9&rp800367a01&14.8&10.5& 7.6& 7.4&  1.2&32&  14&  12&  18& 11.5&  1.4&  1.4& 0.02$^{+0.40}_{-0.56}$&0.93&  2.8$^{ +0.7}_{ -0.6}$&S\\
RXJ033402.54-390048.7&rp800367a01&14.8&13.9& 4.9&12.8&  2.3&50&  34&  33&  50& 25.5&  3.1&  3.2&-0.06$^{+0.30}_{-0.23}$&1.00&  7.2$^{ +1.0}_{ -0.9}$&S\\
RXJ033729.08-253056.2&rp300079n00&41.7&11.6&10.1& 4.7&  1.5&50& 100&  30&  25& 96.9& 15.7&  5.3& 0.03$^{+0.45}_{-0.82}$&0.93&  1.0$^{ +0.2}_{ -0.2}$&I\\
RXJ033736.63-252052.2&rp300079n00&41.7& 4.2& 5.0& 8.8&  1.1&40&  61&  26&  36& 62.8& 10.2&  3.4&-0.51$^{+0.55}_{-0.56}$&1.00&  1.5$^{ +0.3}_{ -0.2}$&S\\
RXJ033827.42-252400.0&rp300079n00&41.7& 7.9& 1.6&26.2&  0.9&54& 110&  80& 143&105.3& 17.1&  5.8&-0.62$^{+0.18}_{-0.27}$&1.00&  6.3$^{ +0.5}_{ -0.5}$&S\\
RXJ034043.77-183457.5&rp600163n00&17.7& 7.5& 3.7&14.4&  1.0&54&  30&  22&  56& 27.6&  9.9& 10.7&-0.68$^{+0.70}_{-1.01}$&0.97&  4.8$^{ +1.1}_{ -0.6}$&I\\
RXJ034110.98-445654.0&rp900495n00&45.0&10.6& 8.0& 6.0&  2.0&29&  27&   8&  25& 27.8&  2.6&  1.9&-1.00$^{+0.86}_{-1.20}$&0.99&  1.2$^{ +0.3}_{ -0.3}$&I\\
RXJ034119.02-441033.3&rp900632n00&42.3& 9.9& 4.1&11.4&  1.7&22&  20&  21&  33& 16.1&  1.8&  1.0& 0.06$^{+0.34}_{-0.34}$&0.96&  2.6$^{ +0.4}_{ -0.4}$&S\\
RXJ034130.14-450309.8&rp900495n00&45.0&11.1& 6.2& 7.6&  1.3&43&  51&  18&  30& 59.6&  5.7&  4.1&-0.47$^{+0.71}_{-0.60}$&0.99&  1.2$^{ +0.2}_{ -0.3}$&I\\
RXJ034248.22-450507.2&rp900495n00&45.0&12.5&14.3& 4.7&  2.0&54&  84&  15&  26& 86.8&  8.3&  6.0&-0.59$^{+0.85}_{-1.40}$&0.97&  0.9$^{ +0.2}_{ -0.2}$&I\\
RXJ043420.48-082136.7&rp700290n00& 6.5&14.1& 7.0&10.7&  2.3&54&  24&  10&  38& 22.3&  3.4&  1.8&-1.17$^{+1.21}_{-0.95}$&0.98& 11.5$^{ +2.7}_{ -1.8}$&S\\
RXJ045304.40-531710.5&rp600436n00&21.8&11.3& 8.5& 5.6&  0.6&50&  37&   7&  21& 32.0&  3.1&  2.8&-0.80$^{+1.07}_{-1.79}$&0.91&  1.6$^{ +0.5}_{ -0.4}$&S\\
RXJ045558.99-753229.1&rp200921n00&22.8&18.0& 1.9&63.7&  8.0&54&  25&  87& 719& 19.6&  6.3&  4.7&-2.43$^{+0.26}_{-0.26}$&1.00& 90.8$^{ +4.9}_{ -3.9}$&S\\
RXJ052839.93-325148.5&rp300004n00& 6.8&10.1& 5.5& 7.9&  1.8&36&   8&  18&  20&  6.1&  2.1&  0.6& 0.09$^{+0.40}_{-0.38}$&0.92&  7.2$^{ +1.4}_{ -1.4}$&S\\
RXJ053018.72-462509.7&rp300334n00&32.1&13.4& 7.2&10.5&  1.8&54&  43&  18&  46& 44.4&  7.4&  6.7&-0.73$^{+0.66}_{-1.10}$&0.98&  2.5$^{ +0.5}_{ -0.4}$&S\\
RXJ053219.84-462550.5&rp300334n00&32.1& 7.8& 4.9&10.2&  1.1&54&  51&  14&  40& 45.1&  7.5&  6.8&-1.08$^{+1.24}_{-1.26}$&0.97&  2.0$^{ +0.5}_{ -0.4}$&I\\
RXJ053242.79-462400.9&rp300334n00&32.1&11.5& 7.7& 6.5&  1.1&54&  51&  12&  35& 42.1&  7.0&  6.3&-1.17$^{+1.11}_{-1.82}$&0.95&  1.8$^{ +0.5}_{ -0.4}$&I\\
RXJ082640.20+263112.3&rp200453n00&13.7& 7.2& 5.2&10.2&  1.0&54&  14&   8&  30& 17.3&  2.3&  2.1&-1.27$^{+0.93}_{-1.06}$&1.00&  3.8$^{ +0.8}_{ -0.9}$&S\\
RXJ085240.51+134651.8&rp700887n00&17.7& 9.2& 1.7&19.1&  2.4&54&  22&  45&  71& 26.1&  3.5&  3.6&-0.02$^{+0.25}_{-0.32}$&0.98&  8.2$^{ +0.9}_{ -0.9}$&S\\
RXJ085340.52+134924.9&rp700887n00&17.7& 8.2& 1.5&28.0&  0.9&40&  22&  44& 109& 14.3&  1.9&  2.0&-0.19$^{+0.23}_{-0.34}$&1.00& 12.0$^{ +1.0}_{ -1.2}$&S\\
RXJ085420.97+135439.6&rp700887n00&17.7&17.4&10.8& 5.5&  0.7&54&  33&   3&  17& 20.4&  2.7&  2.8&-2.07$^{+2.25}_{-2.77}$&0.86&  1.8$^{ +0.9}_{ -0.6}$&S\\
RXJ085851.49+141150.7&rp700436n00&18.7& 3.8& 2.8&11.8&  2.1&36&  15&  10&  42& 14.0&  1.8&  1.6&-1.32$^{+0.82}_{-0.84}$&1.00&  4.0$^{ +0.8}_{ -0.6}$&S\\
RXJ085906.68+140307.2&rp700436n00&18.7& 8.6& 5.9& 7.5&  4.1&54&  20&  10&  27& 31.5&  4.0&  3.6&-1.06$^{+0.85}_{-1.29}$&0.99&  2.4$^{ +0.7}_{ -0.5}$&I\\
RXJ085934.64+141457.6&rp700436n00&18.7&14.3& 6.4&13.5&  0.9&54&  23&  27&  73& 29.8&  3.8&  3.4&-0.65$^{+0.37}_{-0.57}$&1.00&  7.8$^{ +1.1}_{ -0.8}$&I\\
 \end{tabular}
 \end{table*}
 \newpage
 \begin{table*}
 \tabcolsep=1mm
 \begin{tabular}{llllllllllllllcccc}
 \multicolumn{18}{l}{{\bf Table \ref{tab:hardcat}.} continued}\\
 1&2&3&4&5&6&7&8&9&10&11&12&13&14&15&16&17&18\\
                                 Source      & ROR & Exp &Offax & Perr & $\sigma$ & PSF & r & c1 & c2 & c3 & b1 & b2 & b3 & $\alpha$ & P& Flux & Spl\\
 &&(ks)&$(")$&(')&&$\chi^{2}_{/\nu}$&"&&&&&&&&$<0.5$&0.5-2&\\
 &&&&&&&&&&&&&&&&&\\
RXJ090518.27+335006.0&rp700326n00&13.5&17.9& 8.5& 9.2&  2.4&54&  21&   8&  43& 18.2&  3.8&  2.3&-2.43$^{+1.80}_{-1.09}$&1.00&  7.6$^{ +2.3}_{ -1.3}$&S\\
RXJ090923.64+423629.2&rp700329a01&18.8&17.6& 4.0&12.6&  1.6&54&  51&  47&  49& 37.1&  6.4&  3.9& 0.23$^{+0.20}_{-0.22}$&0.93&  7.4$^{ +1.0}_{ -0.9}$&S\\
RXJ090926.82+541429.2&rp201382n00&31.5& 9.5& 5.0&10.7&  5.7&25&  12&  16&  28& 12.5&  1.7&  1.2&-0.13$^{+0.43}_{-0.50}$&0.97&  2.2$^{ +0.4}_{ -0.4}$&I\\
RXJ091006.80+425449.8&rp700329a01&18.8& 6.1& 4.7& 9.6&  1.0&36&  23&  14&  25& 22.6&  3.9&  2.4&-0.40$^{+0.61}_{-0.68}$&0.99&  2.4$^{ +0.6}_{ -0.4}$&I\\
RXJ091049.66+430405.5&rp700329a01&18.8&17.2& 5.3& 9.3&  1.7&54&  45&  32&  44& 44.7&  7.8&  4.7&-0.20$^{+0.34}_{-0.40}$&1.00&  5.7$^{ +0.9}_{ -0.8}$&I\\
RXJ091908.27+745305.6&rp700882n00&10.8&13.6& 9.1& 8.6&  1.2&50&  18&  11&  28& 14.0&  2.4&  2.2&-0.35$^{+0.54}_{-0.82}$&0.98&  4.7$^{ +1.0}_{ -0.9}$&S\\
RXJ092103.75+621504.3&rp700211n00&16.0& 3.8& 4.1&10.2&  1.7&32&   9&  14&  28&  8.6&  1.8&  1.3&-0.02$^{+0.50}_{-0.63}$&0.90&  3.4$^{ +0.6}_{ -0.7}$&S\\
RXJ094144.51+385434.8&rp700387n00&15.8& 7.7& 6.5& 8.5&  1.3&54&  35&  17&  19& 38.7&  4.7&  3.8&-0.12$^{+0.57}_{-0.58}$&0.95&  2.1$^{ +0.5}_{ -0.5}$&S\\
RXJ094356.53+164244.1&rp400141n00& 8.2&12.0& 6.2& 7.4&  0.6&54&  13&   7&  19& 11.0&  2.1&  1.9&-0.44$^{+0.80}_{-1.32}$&0.94&  4.1$^{ +1.2}_{ -0.9}$&S\\
RXJ095029.49+733106.7&rp701214n00& 8.4&17.0&12.2& 6.0&  0.9&54&  11&   7&  20&  9.7&  1.8&  1.7&-0.65$^{+0.70}_{-1.25}$&0.98&  5.3$^{ +1.4}_{ -1.3}$&I\\
RXJ095340.67+074426.1&rp400059n00& 6.5&13.7&11.3& 6.0&  1.4&54&   5&   9&  14&  9.0&  1.5&  1.3&-0.38$^{+0.52}_{-0.94}$&0.96&  4.3$^{ +1.2}_{ -1.1}$&S\\
RXJ100923.93+543635.7&rp900213n00&12.4&11.4& 4.5&12.7&  1.5&54&  64&  21&  44& 38.2&  7.9&  4.0& 0.13$^{+0.27}_{-0.25}$&0.97&  5.9$^{ +1.0}_{ -0.9}$&S\\
RXJ100946.60+553658.3&rp900215n00&15.1& 9.2&11.0& 4.9&  0.9&54&  42&  14&  17& 40.0&  4.2&  3.4&-0.13$^{+0.50}_{-0.68}$&0.96&  1.9$^{ +0.5}_{ -0.5}$&I\\
RXJ101008.53+513334.9&rp700265a01&14.3&11.5&10.7& 6.0&  1.2&50&  34&  17&  17& 36.8&  4.9&  2.7&-0.28$^{+0.57}_{-0.56}$&0.99&  2.3$^{ +0.6}_{ -0.6}$&S\\
RXJ101009.05+525901.5&rp700263n00&11.7&14.0&10.3& 8.3&  2.1&54&  40&  10&  24& 31.4&  3.1&  2.5&-0.17$^{+0.53}_{-0.81}$&0.98&  3.5$^{ +0.9}_{ -0.7}$&S\\
RXJ101028.63+512841.2&rp700265a01&14.3&16.5&16.6& 4.1&  1.1&54&  37&  21&  16& 33.7&  4.5&  2.5& 0.10$^{+0.36}_{-0.41}$&0.93&  3.0$^{ +0.7}_{ -0.7}$&S\\
RXJ101031.05+503458.6&rp900214n00&12.4&10.5&21.5& 4.7&  0.8&54&  41&  13&  21& 33.4&  4.6&  3.1&-0.06$^{+0.44}_{-0.75}$&0.97&  2.8$^{ +0.8}_{ -0.6}$&I\\
RXJ101033.47+533922.5&rp700264n00&13.5& 6.1& 8.5& 6.0&  1.1&36&  15&   9&  14& 13.2&  1.3&  0.9&-0.24$^{+0.56}_{-0.58}$&0.98&  2.0$^{ +0.5}_{ -0.5}$&S\\
RXJ101058.91+534005.8&rp700264n00&13.5& 7.9& 7.3& 6.4&  1.7&32&  13&   9&  17& 12.2&  1.2&  0.9&-0.45$^{+0.57}_{-0.61}$&1.00&  2.5$^{ +0.6}_{ -0.6}$&I\\
RXJ101112.05+554451.3&rp900215n00&15.1& 7.4& 3.5&17.3&  1.3&54&  49&  31&  57& 40.6&  4.3&  3.4&-0.43$^{+0.29}_{-0.29}$&1.00&  7.0$^{ +0.9}_{ -0.9}$&I\\
RXJ101123.17+524912.4&rp700263n00&11.7&11.2& 5.3& 6.7&  1.7&54&  26&   8&  23& 32.2&  3.2&  2.6&-1.41$^{+0.93}_{-1.22}$&1.00&  3.3$^{ +1.0}_{ -0.9}$&S\\
RXJ101147.48+505002.2&rp900214n00&12.4&15.9&60.0& 8.0&  1.4&54&  41&  16&  27& 32.1&  4.5&  3.0&-0.09$^{+0.44}_{-0.47}$&0.99&  4.6$^{ +1.0}_{ -0.8}$&S\\
RXJ101159.42+281407.7&rp000049n00&25.3&13.7& 9.1& 8.8&  1.3&54&  54&  22&  42& 58.2&  7.7&  5.3&-0.38$^{+0.45}_{-0.82}$&0.98&  3.0$^{ +0.6}_{ -0.4}$&S\\
RXJ103133.93-142157.1&rp700461n00&13.1& 7.3& 5.7& 8.6&  1.7&29&   2&   5&  24&  6.1&  1.3&  1.4&-1.41$^{+1.10}_{-1.33}$&0.99&  3.7$^{ +1.0}_{ -0.9}$&S\\
RXJ103230.88-141925.2&rp700461n00&13.1& 9.0& 6.4& 8.8&  0.9&40&   9&   9&  26& 12.5&  2.6&  2.8&-0.62$^{+0.74}_{-1.32}$&0.94&  3.6$^{ +0.8}_{ -0.8}$&S\\
RXJ104648.27+541235.4&rp300158n00&10.9& 7.1& 6.7& 7.6&  1.1&36&  12&   3&  23& 11.6&  1.4&  1.3&-2.61$^{+1.36}_{-1.63}$&1.00&  3.9$^{ +1.2}_{ -1.0}$&S\\
RXJ104723.37+540412.6&rp300158n00&10.9&14.4&13.4& 4.9&  1.9&54&  19&   8&  11& 23.4&  2.7&  2.6&-0.56$^{+0.90}_{-1.09}$&0.98&  1.7$^{ +0.7}_{ -0.6}$&S\\
RXJ110210.03+251418.8&rp300291n00&41.8&11.8& 5.5& 8.8&  0.7&54& 113&  28&  52&105.1& 10.8&  6.1&-0.33$^{+0.53}_{-0.58}$&0.99&  2.1$^{ +0.3}_{ -0.3}$&S\\
RXJ110228.24+250327.1&rp300291n00&41.8& 3.1& 1.1&26.4&  1.4&29&  38&  45& 127& 31.2&  3.2&  1.8&-0.71$^{+0.31}_{-0.25}$&1.00&  5.8$^{ +0.5}_{ -0.5}$&S\\
RXJ110231.12+355342.7&rp200127a01&10.1&13.8&19.4& 4.0&  0.6&54&  19&  10&  14& 24.6&  3.1&  2.5&-0.34$^{+0.69}_{-0.99}$&0.95&  2.5$^{ +0.8}_{ -0.8}$&I\\
RXJ110244.57+250959.4&rp300291n00&41.8& 5.3& 9.6& 4.4&  0.8&43&  62&  17&  20& 74.5&  7.7&  4.4&-0.47$^{+0.96}_{-0.84}$&0.98&  0.8$^{ +0.2}_{ -0.2}$&I\\
RXJ110431.75+355208.5&rp200127a01&10.1&17.1&12.2& 5.1&  1.0&54&  23&  10&  16& 24.3&  3.0&  2.4&-0.21$^{+0.63}_{-1.02}$&0.94&  3.5$^{ +1.1}_{ -0.9}$&S\\
RXJ110452.52+360414.8&rp200127a01&10.1&18.0&11.4& 7.3&  0.9&54&  24&  13&  23& 24.4&  3.0&  2.5&-0.26$^{+0.67}_{-0.63}$&0.97&  5.5$^{ +1.3}_{ -1.1}$&S\\
RXJ110707.63+723634.8&rp700872n00&10.0& 2.8& 9.2& 4.5&  1.2&29&   3&   4&   9&  6.6&  1.0&  1.3&-0.59$^{+0.96}_{-1.44}$&0.93&  1.5$^{ +0.7}_{ -0.5}$&S\\
RXJ110742.05+723236.0&rp700872n00&10.0& 4.4& 2.7&16.1&  0.5&47&  24&  20&  45& 15.8&  2.3&  3.2& 0.12$^{+0.37}_{-0.45}$&0.90&  8.1$^{ +1.2}_{ -1.1}$&S\\
RXJ111750.51+075712.8&rp700358n00&13.6&12.8& 2.1&24.4&  2.4&47&  20&  51& 103& 17.9&  2.8&  2.1&-0.15$^{+0.23}_{-0.29}$&1.00& 16.5$^{ +1.6}_{ -1.4}$&S\\
RXJ111926.34+210646.1&rp700228n00&20.1&13.0& 3.5&12.6&  1.5&54&  76&  47&  58& 48.9&  8.3&  4.6& 0.29$^{+0.18}_{-0.19}$&0.90&  5.9$^{ +0.7}_{ -0.7}$&S\\
RXJ111935.63+133407.4&rp700010n00&13.2&10.1& 5.9& 6.2&  1.2&54&  22&  14&  21& 34.9&  5.5&  3.2&-0.72$^{+0.67}_{-1.03}$&0.99&  2.7$^{ +0.7}_{ -0.7}$&S\\
RXJ111942.16+211518.1&rp700228n00&20.1& 8.5& 4.3&12.0&  2.3&54&  61&  35&  37& 52.5&  8.9&  4.9& 0.13$^{+0.30}_{-0.30}$&0.95&  3.4$^{ +0.6}_{ -0.5}$&S\\
RXJ112056.87+132726.2&rp700010n00&13.2&12.6&12.2& 4.9&  0.8&54&  27&   6&  15& 33.1&  5.2&  3.0&-1.65$^{+1.48}_{-2.12}$&0.99&  1.7$^{ +0.8}_{ -0.6}$&S\\
RXJ112838.96-041628.0&rp700372n00&15.7&12.1& 7.4& 4.6&  3.1&54&  19&   9&  17& 30.9&  4.2&  2.7&-0.63$^{+1.01}_{-1.47}$&0.94&  1.9$^{ +0.6}_{ -0.5}$&S\\
RXJ113624.52+295952.8&rp200091n00&25.1&12.0& 5.9& 7.2&  1.6&47&  35&  12&  33& 35.9&  6.2&  3.3&-1.28$^{+1.03}_{-1.32}$&1.00&  2.3$^{ +0.6}_{ -0.5}$&I\\
RXJ114621.27+285320.6&rp300287n00&15.9&11.0&19.1& 4.6&  1.5&43&  21&  15&  16& 23.1&  2.5&  2.0& 0.05$^{+0.44}_{-0.56}$&0.90&  2.2$^{ +0.5}_{ -0.5}$&S\\
RXJ114654.52+283939.9&rp300287n00&15.9& 4.8& 6.2&10.3&  3.1&54&  49&  14&  35& 39.1&  4.3&  3.4&-0.19$^{+0.69}_{-0.73}$&0.97&  3.6$^{ +0.8}_{ -0.6}$&I\\
RXJ114708.66+283001.2&rp300287n00&15.9&14.9&12.7& 6.9&  1.6&47&  21&  12&  22& 25.0&  2.7&  2.2&-0.47$^{+0.74}_{-0.63}$&0.99&  3.0$^{ +0.7}_{ -0.6}$&S\\
RXJ114736.94+284131.4&rp300287n00&15.9&10.6& 4.9& 8.1&  0.9&54&  37&  13&  22& 38.0&  4.2&  3.3&-0.30$^{+0.63}_{-0.85}$&0.97&  2.4$^{ +0.7}_{ -0.5}$&S\\
RXJ115952.10+553212.1&rp700055n00&44.7&17.2& 1.4&73.6&  6.1&54& 418& 410& 698&107.0& 15.9&  9.7& 0.17$^{+0.05}_{-0.04}$&1.00& 40.1$^{ +1.5}_{ -1.2}$&S\\
RXJ120403.79+280711.2&rp700232n00&12.5&15.8& 1.1&38.1& 12.6&54& 156& 123& 246& 36.4& 11.1&  4.5& 0.38$^{+0.09}_{-0.07}$&0.95& 44.9$^{ +2.9}_{ -2.2}$&S\\
RXJ120515.93-033424.6&rp201367m01&45.3&13.8& 6.7&12.9&  1.7&54&  98&  33&  79& 91.2& 11.9&  6.7&-0.64$^{+0.53}_{-0.61}$&1.00&  3.2$^{ +0.4}_{ -0.4}$&I\\
RXJ121017.25+391822.6&rp700277n00&19.0& 6.8& 5.9& 8.9&  1.7&54&  82&  37&  52& 77.2& 13.3&  7.0&-0.02$^{+0.35}_{-0.49}$&0.97&  4.6$^{ +0.7}_{ -0.7}$&S\\
RXJ121115.30+391146.8&rp700277n00&19.0&15.4& 4.4&20.9&  1.7&40&  52&  81&  87& 36.7&  6.3&  3.3& 0.27$^{+0.15}_{-0.17}$&0.95& 15.7$^{ +1.3}_{ -1.5}$&S\\
RXJ121803.82+470854.6&rp600546n00&19.9&13.3&12.4& 4.3&  2.2&54&  45&  12&  18& 53.3&  4.3&  3.8&-0.52$^{+0.60}_{-1.07}$&0.99&  1.5$^{ +0.4}_{ -0.4}$&S\\
RXJ121834.72+300957.2&rp700221n00&18.6&10.0& 5.1& 7.7&  0.9&32&  13&  15&  18& 17.7&  3.3&  2.3&-0.19$^{+0.45}_{-0.68}$&0.96&  2.1$^{ +0.5}_{ -0.5}$&S\\
RXJ124913.86-055906.2&rp600262a02&35.2&14.6& 9.0& 6.9&  0.6&43&  51&  30&  41& 51.0& 12.4&  7.1&-0.09$^{+0.57}_{-0.55}$&0.96&  2.4$^{ +0.5}_{ -0.4}$&S\\
RXJ124919.66-060430.2&rp700375n00&10.1& 5.3& 5.2& 7.6&  0.9&50&  24&  21&  23& 27.7&  7.4&  3.6&-0.20$^{+0.63}_{-0.60}$&0.97&  4.0$^{ +1.0}_{ -0.8}$&S\\
RXJ124949.92+405305.2&rp600050n00&73.0&17.7& 7.6& 8.8&  1.7&54& 361& 198& 124&341.6&133.1& 25.0& 0.08$^{+0.26}_{-0.35}$&0.98&  3.5$^{ +0.4}_{ -0.4}$&S\\
 \end{tabular}
 \end{table*}
 \newpage
 \begin{table*}
 \tabcolsep=1mm
 \begin{tabular}{llllllllllllllcccc}
 \multicolumn{18}{l}{{\bf Table \ref{tab:hardcat}.} continued}\\
 1&2&3&4&5&6&7&8&9&10&11&12&13&14&15&16&17&18\\
                                 Source      & ROR & Exp &Offax & Perr & $\sigma$ & PSF & r & c1 & c2 & c3 & b1 & b2 & b3 & $\alpha$ & P& Flux & Spl\\
 &&(ks)&$(")$&(')&&$\chi^{2}_{/\nu}$&"&&&&&&&&$<0.5$&0.5-2&\\
 &&&&&&&&&&&&&&&&&\\
RXJ125447.96+563842.3&rp700208n00&18.2&17.9&16.6& 5.1&  0.6&54&  68&  25&  42& 60.3& 11.4&  5.9&-0.23$^{+0.41}_{-0.76}$&0.99&  5.0$^{ +1.0}_{ -0.8}$&I\\
RXJ125556.56+565850.9&rp700208n00&18.2& 7.1& 8.8& 4.8&  0.7&54&  58&  23&  22& 73.2& 13.8&  7.2&-0.69$^{+0.93}_{-1.31}$&0.99&  1.6$^{ +0.6}_{ -0.5}$&S\\
RXJ131047.80+322958.1&rp700216a00& 7.7& 9.8& 3.4&12.0&  1.0&54&  25&  16&  28& 21.4&  3.2&  2.2&-0.32$^{+0.50}_{-0.44}$&1.00&  6.5$^{ +1.3}_{ -1.2}$&S\\
RXJ131635.62+285942.7&rp100308n00&19.9& 6.8& 1.8&28.2& 10.7&54& 158&  92& 143& 54.7&  7.2&  3.9& 0.35$^{+0.10}_{-0.08}$&0.96& 13.7$^{ +1.0}_{ -1.0}$&S\\
RXJ131646.89+285721.1&rp100308n00&19.9&10.0& 4.6&11.8&  3.9&54&  71&  26&  38& 55.6&  7.3&  4.0& 0.11$^{+0.28}_{-0.38}$&0.96&  3.3$^{ +0.6}_{ -0.5}$&S\\
RXJ131651.64+291251.1&rp100308n00&19.9& 9.1& 3.0&13.0&  1.2&54&  72&  21&  37& 54.2&  7.1&  3.9& 0.13$^{+0.30}_{-0.45}$&0.94&  3.0$^{ +0.5}_{ -0.5}$&I\\
RXJ131653.87+291708.8&rp100308n00&19.9&12.9& 6.7& 8.3&  1.0&54&  52&  15&  16& 56.4&  7.4&  4.0&-0.31$^{+0.67}_{-1.29}$&0.96&  1.2$^{ +0.4}_{ -0.4}$&I\\
RXJ131721.79+291119.1&rp100308n00&19.9&13.7&14.3& 6.8&  1.4&54&  56&  18&  28& 55.5&  7.3&  4.0&-0.40$^{+0.48}_{-0.88}$&0.99&  2.4$^{ +0.6}_{ -0.5}$&I\\
RXJ133146.37+111420.4&rp701034n00& 9.4&12.0&14.5& 5.0&  0.6&50&  21&   8&  23& 21.5&  4.9&  3.5&-1.47$^{+1.30}_{-1.61}$&1.00&  4.0$^{ +1.2}_{ -1.2}$&S\\
RXJ133147.00+105653.0&rp701034n00& 9.4&13.3&13.0& 5.5&  1.4&54&  21&  18&  16& 23.6&  5.4&  3.8& 0.07$^{+0.42}_{-0.75}$&0.88&  3.2$^{ +0.9}_{ -0.9}$&S\\
RXJ133152.51+111643.5&rp701034n00& 9.4&12.7& 1.0&36.5&  1.0&54&  27&  57& 273& 23.7&  5.4&  3.8&-1.71$^{+0.27}_{-0.33}$&1.00& 57.0$^{ +4.1}_{ -3.9}$&S\\
RXJ133948.81+482111.5&rp700473n00& 9.5&15.9&12.9& 7.1&  0.9&54&  19&  12&  23& 20.0&  2.7&  2.1&-0.48$^{+0.68}_{-0.65}$&0.99&  5.3$^{ +1.2}_{ -1.2}$&I\\
RXJ134335.29-001643.7&rp701000a01&24.7& 2.6& 4.0& 7.2&  0.6&54&  45&  21&  19& 57.9&  9.5&  6.0&-0.19$^{+0.57}_{-1.07}$&0.94&  1.1$^{ +0.4}_{ -0.3}$&I\\
RXJ134905.63+600037.4&rp600270n00&35.7&10.8& 5.8& 6.2&  1.6&54&  84&  14&  26& 82.7&  7.6&  4.4&-0.61$^{+0.94}_{-1.37}$&0.97&  1.2$^{ +0.3}_{ -0.3}$&I\\
RXJ135105.69+601538.5&rp600270n00&35.7&14.5& 5.6&12.4&  2.2&54&  78&  44&  65& 80.0&  7.4&  4.3&-0.12$^{+0.33}_{-0.27}$&1.00&  3.8$^{ +0.4}_{ -0.4}$&S\\
RXJ135529.59+182413.6&rp700392n00&11.0& 8.0& 6.2& 8.5&  0.6&54&  28&  16&  25& 27.8&  7.5&  4.2&-0.41$^{+0.84}_{-0.94}$&0.98&  3.6$^{ +0.9}_{ -0.9}$&S\\
RXJ140134.94+542029.2&rp600108n00&21.3&13.8& 9.8& 6.5&  1.6&54&  89&  34&  39& 87.5& 15.8&  9.1&-0.13$^{+0.46}_{-0.59}$&0.98&  2.9$^{ +0.6}_{ -0.6}$&S\\
RXJ140416.61+541618.2&rp600108n00&21.3&10.6& 4.3& 8.7&  1.3&54&  90&  40&  37& 92.7& 16.8&  9.7&-0.07$^{+0.34}_{-0.56}$&0.98&  2.8$^{ +0.6}_{ -0.5}$&S\\
RXJ142754.71+330007.0&rp200329n00&23.9&14.7&18.1& 7.6&  2.0&54&  83&  26&  38& 65.9&  7.7&  4.2& 0.21$^{+0.27}_{-0.39}$&0.90&  3.0$^{ +0.5}_{ -0.5}$&S\\
RXJ150132.45-082507.2&rp200510n00&11.3&10.5& 2.2&17.0&  1.2&54&  23&  27&  92& 23.2& 14.4&  6.5&-1.78$^{+0.87}_{-0.94}$&1.00& 15.7$^{ +2.6}_{ -2.0}$&S\\
RXJ163054.25+781105.1&rp170154n00&26.0& 7.5& 1.5&25.5&  2.0&25&  21&  40& 106& 12.5&  2.6&  1.2&-0.01$^{+0.37}_{-0.29}$&0.98&  9.0$^{ +0.7}_{ -0.9}$&S\\
RXJ163308.57+570258.7&rp200721n00&37.8&11.9& 7.3& 8.8&  1.6&54& 131&  30&  42&134.6& 11.9&  7.3&-0.11$^{+0.42}_{-0.69}$&0.95&  1.9$^{ +0.3}_{ -0.3}$&S\\
RXJ170041.60+641259.0&rp701457n00&20.5& 2.2& 2.2&36.4& 23.6&47&  95& 145& 282& 32.7&  3.3&  3.2& 0.17$^{+0.11}_{-0.10}$&1.00& 27.0$^{ +1.3}_{ -1.5}$&S\\
RXJ170044.36+520545.6&rp700123n00& 6.6&17.7& 8.9& 6.4&  1.2&54&  13&   8&  17& 11.8&  2.0&  1.3&-0.20$^{+0.61}_{-1.13}$&0.93&  6.1$^{ +1.8}_{ -1.3}$&S\\
RXJ170123.32+641413.0&rp701457n00&20.5& 3.4& 2.0&27.0& 18.4&43&  69&  93& 172& 27.9&  2.8&  2.7& 0.24$^{+0.13}_{-0.13}$&0.99& 16.6$^{ +1.2}_{ -1.0}$&S\\
RXJ171112.30+710924.7&rp700875n00&24.1&12.2&11.7& 4.9&  1.1&50&  33&  14&  28& 36.5&  6.9&  3.7&-0.62$^{+0.80}_{-1.41}$&0.95&  2.1$^{ +0.6}_{ -0.4}$&S\\
RXJ204640.48-363147.5&rp201374a01&25.3&13.6& 4.1&15.6&  2.3&54&  57&  48&  82& 63.8& 13.4&  7.2&-0.14$^{+0.45}_{-0.36}$&0.98&  6.3$^{ +0.7}_{ -0.7}$&S\\
RXJ204716.74-364715.1&rp201374a01&25.3&13.2& 3.8&19.7&  2.3&54&  69&  48& 123& 64.1& 13.5&  7.3&-0.64$^{+0.35}_{-0.53}$&1.00&  9.1$^{ +0.9}_{ -0.9}$&S\\
RXJ213807.61-423614.3&rp701180a01& 2.2&14.1& 7.2& 8.3&  0.5&54&   5&   6&  19&  5.8&  1.3&  0.7&-1.10$^{+0.92}_{-1.06}$&1.00& 16.7$^{ +5.1}_{ -3.6}$&S\\
RXJ223619.89-261426.2&rp700873n00&22.9&13.8&11.9& 4.7&  1.0&54&  62&  19&  28& 72.0&  8.4&  5.3&-0.61$^{+0.61}_{-0.96}$&0.99&  2.1$^{ +0.4}_{ -0.6}$&S\\
RXJ223654.00-261230.6&rp700873n00&22.9&18.1&11.2& 5.1&  0.6&54&  45&  17&  41& 56.3&  6.6&  4.2&-1.24$^{+0.74}_{-0.80}$&1.00&  4.1$^{ +0.9}_{ -0.7}$&S\\
RXJ225018.97+242750.8&rp201552n00&15.8& 9.2& 4.1& 8.2&  2.1&54&   3&  12&  26& 11.2&  2.9&  2.8&-0.59$^{+0.62}_{-0.91}$&0.98&  2.9$^{ +0.7}_{ -0.5}$&S\\
RXJ230248.28+084348.1&rp700423n00&16.6&10.6& 2.8&10.9&  1.9&50&  19&  33&  58& 22.1&  3.8&  4.8& 0.13$^{+0.32}_{-0.46}$&0.88&  7.0$^{ +1.0}_{ -0.8}$&I\\
RXJ232505.73+234056.5&rp200322n00&22.6&17.3&12.5& 5.7&  1.2&54&  26&  16&  40& 26.5&  8.8&  8.0&-0.88$^{+0.99}_{-1.36}$&0.97&  3.6$^{ +0.8}_{ -0.8}$&I\\
RXJ235113.89+201347.3&rp701217a01& 5.6& 7.4& 2.2&28.1&  1.8&54&   9&  31&  91&  5.5&  1.5&  1.5&-0.55$^{+0.29}_{-0.44}$&1.00& 30.3$^{ +3.2}_{ -3.1}$&S\\
\vspace{3mm}\\
\multicolumn{18}{l}{\bf Description of columns in Table \ref{tab:hardcat} (see
Sec \ref{sec:method})}\\
\multicolumn{18}{l}{1)\ Source name based on J2000 coordinates of \ros\ source
from PSS}\\
\multicolumn{18}{l}{2)\ \ros\ observation in which source was detected}\\
\multicolumn{18}{l}{3)\ \ros\ PSPC exposure time in kiloseconds after good time screening}\\
\multicolumn{18}{l}{4)\ Offaxis angle of source in arcminutes}\\
\multicolumn{18}{l}{5)\ Positional uncertainty of source in arcseconds 
from PSS}\\
\multicolumn{18}{l}{6)\ PSS significance of source above background level in
units of $\sigma$}\\
\multicolumn{18}{l}{7)\ $\chi^{2}_{/\nu}$ from a fit of the model 
PSPC point spread
function to the source radial profile in 5 arcsecond bins out to 1 arcminute}\\
\multicolumn{18}{l}{8)\ Source circle radius in arcseconds for extraction of 3
colour spectrum}\\
\multicolumn{18}{l}{9)\ Number of counts in PSPC channels 11-41 within the source circle}\\
\multicolumn{18}{l}{10)\ Number of counts in PSPC channels 52-90 within the source circle}\\
\multicolumn{18}{l}{11)\ Number of counts in PSPC channels 91-201 within the source circle}\\
\multicolumn{18}{l}{12)\ Expected number of background counts in PSPC channels
11-41 within the source circle}\\
\multicolumn{18}{l}{13)\ Expected number of background counts in PSPC channels
52-90 within the source circle}\\
\multicolumn{18}{l}{14)\ Expected number of background counts in PSPC channels
91-201 within the source circle}\\
\multicolumn{18}{l}{15)\ Fitted power law spectral slope with 68\% uncertainty}\\
\multicolumn{18}{l}{16)\ Probability of source having spectral slope $\alpha$ $<$
(harder than) 0.5}\\
\multicolumn{18}{l}{17)\ 0.5 - 2 keV flux and 68\% uncertainty in units of 
$10^{-14}$ \ecs }\\
\multicolumn{18}{l}{18)\ Optical counterpart flag: `S' means spectroscopic
sample, `I' means imaging sample}\\
\end{tabular}
\end{table*}

\end{document}